\documentclass[sigconf,natbib=true]{acmart}
\usepackage[utf8]{inputenc}
\usepackage{amsmath}
\usepackage{amsfonts}
\usepackage{pdfpages}
\usepackage{subcaption}
\usepackage{epstopdf}
\usepackage{epsfig}
\usepackage{graphicx}
\usepackage{booktabs}
\usepackage{multirow}
\usepackage{comment}

\usepackage{algorithm}
\usepackage{algorithmic}

\usepackage{tabularx}

\usepackage{pgfgantt}
\usepackage{lscape}
\usepackage{ctable, booktabs, makecell}
\usepackage{wrapfig, layouts}
\usepackage{inputenc}
\usepackage{geometry}

\newtheorem{assumption}{Assumption}
\newtheorem{theorem}{Theorem}

\AtBeginDocument{%
  \providecommand\BibTeX{{%
    \normalfont B\kern-0.5em{\scshape i\kern-0.25em b}\kern-0.8em\TeX}}}

\setcopyright{acmlicensed}
\copyrightyear{2024}
\acmYear{2024}
\setcopyright{acmlicensed}\acmConference[CIKM '24]{Proceedings of the 33rd ACM International Conference on Information and Knowledge Management}{October 21--25, 2024}{Boise, ID, USA}
\acmBooktitle{Proceedings of the 33rd ACM International Conference on Information and Knowledge Management (CIKM '24), October 21--25, 2024, Boise, ID, USA}
\acmDOI{10.1145/3627673.3679633}
\acmISBN{979-8-4007-0436-9/24/10}

\newcommand{\MYhref}[3][blue]{\href{#2}{\color{#1}{#3}}}%

\begin{document}

\title{ROLeR: Effective Reward Shaping in Offline Reinforcement Learning for Recommender Systems}

\author{Yi Zhang}
\email{uqyzha91@uq.edu.au}
\affiliation{\institution{The University of Queensland \\ CSIRO DATA61}
  \city{Brisbane}
  \country{Australia}}
\author{Ruihong Qiu}
\email{r.qiu@uq.edu.au}
\affiliation{\institution{The University of Queensland}
  \city{Brisbane}
  \country{Australia}}
\author{Jiajun Liu}
\email{jiajun.liu@csiro.au}
\affiliation{\institution{CSIRO DATA61 \\ The University of Queensland}
  \city{Brisbane}
  \country{Australia}}
\author{Sen Wang}
\email{sen.wang@uq.edu.au}
\affiliation{\institution{The University of Queensland}
  \city{Brisbane}
  \country{Australia}}

\renewcommand{\shortauthors}{Yi Zhang, Ruihong Qiu, Jiajun Liu and Sen Wang}

\begin{abstract}

Offline reinforcement learning (RL) is an effective tool for real-world recommender systems with its capacity to model the dynamic interest of users and its interactive nature. 
Most existing offline RL recommender systems focus on model-based RL through learning a world model from offline data and building the recommendation policy by interacting with this model.
Although these methods have made progress in the recommendation performance, the effectiveness of model-based offline RL methods is often constrained by the accuracy of the estimation of the reward model and the model uncertainties, primarily due to the extreme discrepancy between offline logged data and real-world data in user interactions with online platforms. 
To fill this gap, a more accurate reward model and uncertainty estimation are needed for the model-based RL methods.
In this paper, a novel model-based \textbf{R}eward Shaping in \textbf{O}ffline Reinforcement \textbf{Le}arning for \textbf{R}ecommender Systems, \textbf{ROLeR}, is proposed for reward and uncertainty estimation in recommendation systems.
Specifically, a non-parametric reward shaping method is designed to refine the reward model. In addition, a flexible and more representative uncertainty penalty is designed to fit the needs of recommendation systems. Extensive experiments conducted on four benchmark datasets showcase that ROLeR achieves state-of-the-art performance compared with existing baselines. Source code can be downloaded at \MYhref{https://github.com/ArronDZhang/ROLeR}{this address}.

\end{abstract}



\maketitle

\section{Introduction}

Recommender systems (RS) serve as a forever-running engine on online platforms with large volumes of data to create personalised experiences for massive users through services like content recommendations and targeted advertising~\cite{bobadilla2013rs_survey,burke2002hybridRS}. The performance of the recommender systems significantly bolsters a company's market competitiveness, especially in industries where user engagement and retention are essential to the business.

To dynamically model user preference, over the past few years, reinforcement learning (RL)~\cite{sutton2018reinforcement} has been incorporated into recommender systems by modeling the entire recommendation process as an interactive problem~\cite{afsar2022rsrl}. In this process, interactions between users and the recommender system are formulated as a sequence of states, actions, and rewards within an environment. The RL-based approaches can continuously learn and adapt from each user interaction, being able to adjust quickly to changing user preferences and behaviors~\cite{hong2020nonintrusive}, optimizing users’ long-term engagement~\cite{wang2022surrogate}. Despite the effectiveness of RL, it is unrealistic for an uncompleted recommender system to perform expensive online interactions with users to collect training data~\cite{gao2023dorl,gao2023cirs,chen2023deep,chen2023opportunities}.

\begin{figure}[!t]
\centering
\includegraphics[trim=0cm 0cm 0cm 0cm, clip, width=0.98\columnwidth]{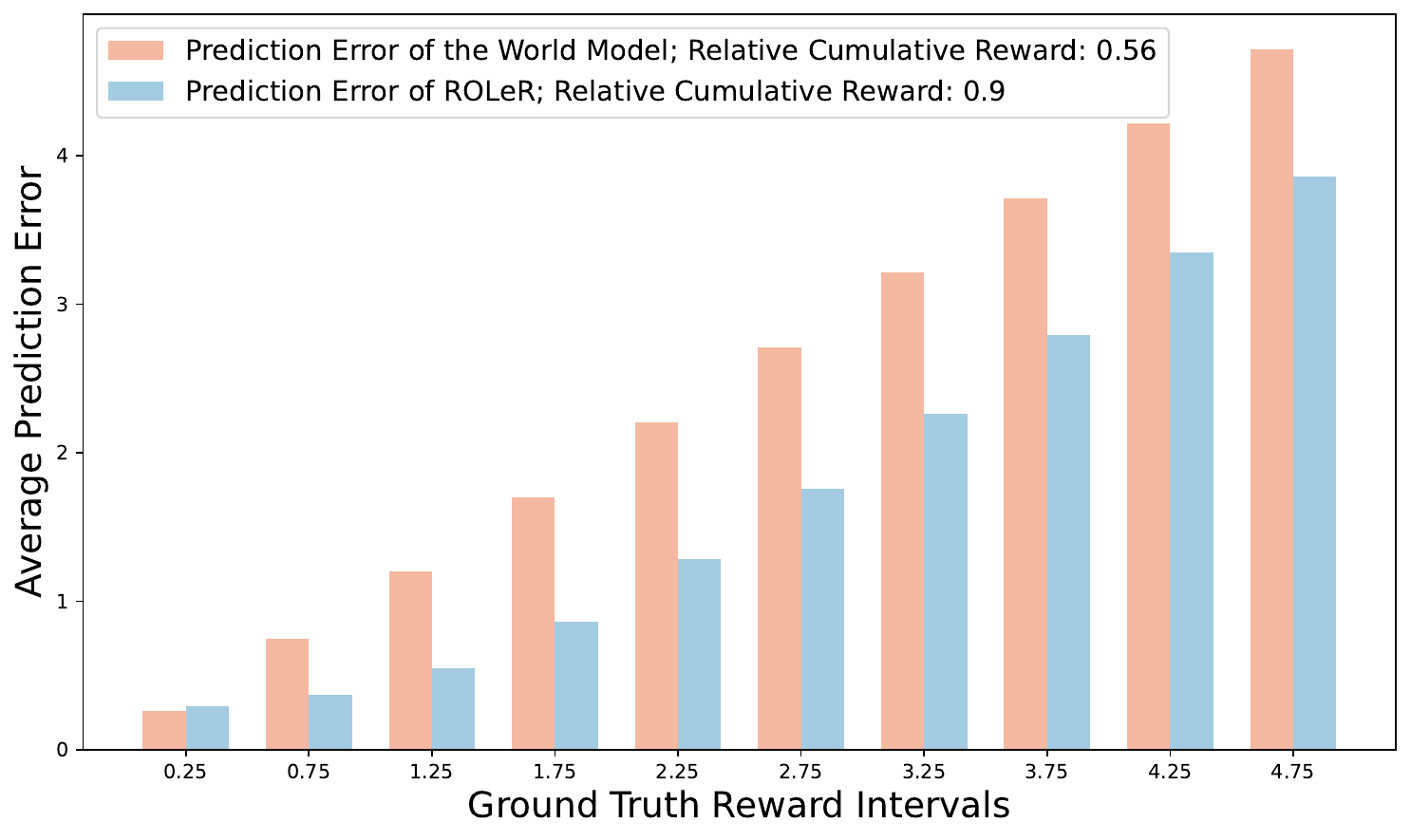}
\vspace{-1.7em}
\caption{The reward estimation error of a world model and ROLeR across different intervals. Our training-free reward shaping constantly outperforms that of the current world model, reaching a higher relative cumulative reward.}
\label{fig:main_result_cossim}
\end{figure}

Although these methods have made progress in the recommendation performance, the effectiveness of RL methods is often constrained by the accuracy of the estimation of the reward model and the model uncertainties. This is because of the extreme discrepancy between offline data and real-world data on online platforms. Recent methods ~\cite{gao2023dorl,zheng2021dgcn} target at the Matthew Effect logged in the offline data with an entropy penalty to enhance the recommendation diversity. Others focus on the individual perspective for the filter bubble issue~\cite{gao2023cirs,liu2021interaction} in user preference. The efficiency of these model-based offline reinforcement learning (RL) methods is frequently limited by the precision of the reward model estimation and uncertainty in the reward model, which are predominantly attributed to the significant disparity between offline logged data and real-world data in user interactions on online platforms. Revisiting the literature of offline RL for RS, it is worth noting that neither behaving conservatively nor encouraging exploration can empower the policy to learn from offline data under the influence of an inaccurate reward estimation. In Figure \ref{fig:main_result_cossim}, we demonstrate the influence of the accuracy of the reward functions on a benchmark dataset -- KuaiRec~\cite{gao2022kuairec}. The reward prediction error of the state-of-the-art method, DORL~\cite{gao2023dorl}, is generally high in all reward intervals in the red bar. An ideal situation would require the reward estimation error to be smaller for an effective RS development.

To address the aforementioned problems, we propose a novel model-based \textbf{R}eward Reshaping in \textbf{O}ffline Reinforcement \textbf{Le}arning method for \textbf{R}ecommender Systems, \textbf{ROLeR} to improve the reward function during policy learning and decouple the uncertainty penalty with the ensemble of the world model. In RS, although the users' tastes may vary from person to person and change over time, finding similar users who share analogous interests is feasible. Based on this intuition, we try to discover the patterns within the user-item interactions from offline data. We find that both the users' historical interaction data and the learned user embedding from the world model can be used as user indicator features. And if the indicator features of one user match that of another user, their feedback on a certain group of items can be mutually inferred. Thus, we regard a user's indicator feature as a soft label to retrieve its nearest neighbors. Then, the user's feedback on a certain item can be inferred by that from these neighbors. In other words, our reward shaping method is a non-parametric clustering-based one, simple yet effective. The accuracy of the reward function can be substantially improved.
On the other hand, to cooperate with our reward shaping module, we develop an uncertainty estimation method based on the quality of the clustering. The distance between a user and its nearest neighbors is used as the uncertainty, measuring the utility of current reward estimation. Therefore, this uncertainty penalty works as the complementary for the reward shaping part, releasing the need of an ensemble of world models.
Our contributions are summarised as:
\vspace{-0.5em}
\begin{itemize}
\item We find out the accuracy of the reward function prediction substantially determines the RS performance and propose a novel reward shaping method that is non-parametric for model-based offline RL for RS.
\item We develop a new uncertainty penalty that integrates with the reward shaping part for better generalization capacity and does not rely on an ensemble of world models. 
\item We demonstrate the effectiveness of the proposed ROLeR on four recommendation benchmarks, KuaiRand, KuaiEnv, Coat and Yahoo, to showcase the superior performance among the state-of-the-art methods and strong baselines.
\end{itemize}
\vspace{-0.5em}

\section{Related Work}
\vspace{-0.2em}
\subsection{RL in Recommendation Systems}
Since supervised RSs have difficulties in capturing the dynamics of user preference~\cite{chen2023deep,zhang2019deep,qiu2019rethinking,qiu2022contrastive,qiu2020gag,qiu2021memory}, more deep RL~\cite{sutton2018reinforcement} methods are employed. Some directions~\cite{zhao2018deep,xiao2020deep} focuses on formulating the RS to a Markov Decision Process (MDP), investigating the state representation~\cite{huang2022state_repr,lei2019social}, user profiling~\cite{user_profile1,user_profile12}, and action representation~\cite{GCQN_RS}. Further, based on the transition modeling, the literature can be divided into model-based methods and model-free methods. Model-based RL~\cite{yu2020mopo,yu2021combo} utilizes offline data to build the transition model. Further, model-free RL~\cite{kumar2020cql,fujimoto2019bcq} gains more attention especially when an online interaction environment is available, including Double DQN~\cite{van2016ddqn} to model the user dynamics~\cite{zheng2018drn} and utilizing a multi-agent setting~\cite{lowe2017maddpg} to tackle the sub-optimum issue~\cite{zhao2020deepchain}.
\vspace{-0.4em}

\subsection{Offline Reinforcement Learning}
Recently, offline RL has attracted great attention from the research community~\cite{levine2020offline,prudencio2023survey}. The gap between the learned policy and the real data results in the overestimation issue~\cite{wang2020statistical, kumar2019stabilizing}. To solve this problem, model-free methods regulate the policy learning to be conservative. For instance, BCQ~\cite{fujimoto2019bcq} avoid using the out-of-distribution (OOD) experience in policy iteration. Besides, CQL~\cite{kumar2020cql} constrains the usage of OOD data in the iteration of the state-action value function. CRR~\cite{wang2020crr} limits the policy improvement conditioning on the discrepancy between the behavior policy and the learning policy. Though they achieve high accuracy in value estimation, they are also limited in policy improvement~\cite{xu2022policy}. To deal with the extrapolate error introduced by the world model, recent methods such as MOReL~\cite{kidambi2020morel} and MOPO~\cite{yu2020mopo} use the distances calculated through an ensemble of world models to penalize the overestimated values. COMBO~\cite{yu2021combo} penalizes the rewards that tend to be OOD and controls the degree of pessimism based on a sampling distribution.

\section{Preliminaries}
\vspace{-0.2em}
\subsection{Interactive Recommendation}
The interactive recommendation is a comprehensive simulation of real-world recommendations where the systems need to continuously recommend items to users.
\vspace{-0.4em}

\subsection{Reinforcement Learning Formulation}
Reinforcement Learning (RL) aims to make a sequence of decisions to maximize the long-term return. RL problems can be formulated as a Markov Decision Process (MDP) with a tuple $G = <\mathbf{S},\mathbf{A},T,r,\gamma>$. $\mathbf{S}$ is the state space and each $\mathbf{s} \in \mathbf{S}$ refers to a specific state. $\mathbf{A}$ is the action space consisting of all the potential actions. $T$ stands for the state transition of the environment as $T\{\mathbf{s_t}, \mathbf{a_t}, \mathbf{s_{t+1}}\} = P(\mathbf{s_{t+1}}|\mathbf{s}=\mathbf{s_t},\mathbf{a}=\mathbf{a_t})$. It describes the dynamics of the environment. $R$ is the reward function as $r_t = R(\mathbf{s_t}, \mathbf{a_t})$, representing the reward of taking action $\mathbf{a_t}$ at state $\mathbf{s_t}$. At last, $\gamma$ is the discount factor used to balance the current reward and future return. The objective of RL is to learn a policy $\pi$ which can maximize the long-term return: $G_t = \sum_{\mathbf{s}=\mathbf{s_t}}^T \gamma^t r(\mathbf{s_t}, \mathbf{a_t})$, where $\mathbf{s_t}$ stands for the start state, $T$ is the last state and t is the timestep. 
In addition, the state value function and state-action value function of a given policy $\pi$ are $V_{\pi}(\mathbf{s}) = \mathbf{E}_{\pi}[G_t|\mathbf{s}=\mathbf{s_t}]$ and $Q_{\pi}(\mathbf{s}, \mathbf{a}) = \mathbf{E}_{\pi}[G_t|\mathbf{s}=\mathbf{s_t}, \mathbf{a}=\mathbf{a_t}]$, respectively. In finite MDP, where the state space and action space are finite, an optimal policy, $\pi_*$, whose expected return is no less than that of all other policies. Theoretically, it can be learned through policy iteration and value iteration with the Bellman Equation:
\begin{equation}
\abovedisplayskip=2pt
\belowdisplayskip=0pt
\label{v_s}
    V_{k+1}(\mathbf{s}) = \max_\mathbf{a} \sum_{\mathbf{s'},r}P(\mathbf{s'}, r|\mathbf{s},\mathbf{a})[r + \gamma V_k(\mathbf{s'})],
\end{equation}
\begin{equation}
\abovedisplayskip=0pt
\belowdisplayskip=0pt
\label{q_sa}
    Q_{k+1}(\mathbf{s}, \mathbf{a}) =  \sum_{\mathbf{s'},r}P(\mathbf{s'}, r|\mathbf{s},\mathbf{a})[r + \gamma \max_{\mathbf{a'}} Q_k(\mathbf{s'},\mathbf{a'})],
\end{equation}
where $k$ is the iteration index and we use $\mathcal{T}$ to represent the Bellman Operator in the following paragraphs.
\vspace{-0.3em}

\subsection{Offline Reinforcement Learning}
Compared to expensive online interaction, offline data are usually abundant and easily accessible. Therefore, offline RL, which investigates how to utilize offline datasets to train policies, attracts increasing attention in current research due to its potential efficiency. The offline dataset $D = \{(\mathbf{s_t}, \mathbf{a_t}, \mathbf{s_{t+1}}, r_t)\}$ is collected by one or more behavior policies $\pi_{B_i}$. One intuitive challenge of offline RL is aroused by the inevitable gap between the distribution of offline data and the evaluated environment. It is difficult for the learning policy to estimate the value function on rarely seen and even unseen states. Directly using the online RL methodologies tends to overestimate the value functions on these states due to the maximization over available actions during decision making. The corresponding solutions can be categorized into two classes. The model-free offline RL adds constraints on the behavior policies and the learning policy to avoid risky decisions. But it limits the generation of offline RL. The model-based offline RL simulates the environments based on the offline datasets to enable the learning agent to interact with them. Then, they introduce the uncertainty estimation as a penalty in the reward function to encourage conservative actions by $r = \hat{r} - \lambda_U P_U$. This branch suits the RS since the offline interactions are highly sparse. Diving into model-based offline RL for RS, it first models both the transition function and reward function to obtain a world model denoted as $G' = <\mathbf{S},\mathbf{A},\hat{T},\hat{R},\gamma>$, where $\hat{T}$ and $\hat{R}$ are the estimated transition and reward functions.

In addition, many recent efforts in this domain rely on an ensemble of world models to calculate uncertainty. It unnecessarily binds the world model learning and uncertainty penalty.

\section{Method}
\vspace{-0.2em}
\subsection{Problem Definition}
\vspace{-0.2em}
Considering reinforcement learning for the recommendation system, we demonstrate the problem formulation in this part. Recalling the MDP tuple $G = <\mathbf{S},\mathbf{A},T,r,\gamma>$, each $\mathbf{s} \in \mathbf{S}$ corresponds to a user state which usually consists of the users' side information such as personal interests and dynamic features like the recent interaction history. The action space $\mathbf{A}$ is the item set and each action $\mathbf{a}$ corresponds to one action $\mathbf{a_t}$. The reward $r(s_t, a_t)$ is from the feedback of recommending item $\mathbf{a_t}$ at state $\mathbf{s_t}$. It depends on the specific datasets, \textit{for instance}, the reward can be the watched ratio in a short video platform or ratings for a film forum. The transition function here is special as it is also a state tracker used for encoding states autoregressively: $\mathbf{s'} = f(\mathbf{s},\mathbf{a},r)$. So it can be implemented using sequential models. At last, the ultimate goal of the recommendation system is to learn a policy $\pi$ which can maximize the cumulative user experience: $\arg \max_{\pi} \mathbf{E}_{\tau \sim \pi}[\sum_{(\mathbf{s},\mathbf{a}) \in \tau} \gamma^t r(\mathbf{s}, \mathbf{a})]$, where $\tau$ is a trajectory sampled with policy $\pi$. In this paper, we focus on the model-based offline RL. And we follow a state-of-the-art method -- DORL to completely illustrate the whole learning process. It usually consists of two stages: world model learning with offline interaction history and train the recommendation policy on this environment. 
\vspace{-0.5em}

\subsection{World Model Learning}
\vspace{-0.2em}
In this part, we mainly focus on the simulation of an environment. The input for this stage is the offline history. They are used as the training set of a supervised prediction model. In DORL and CIRS, they use DeepFM. The outputs are the item embedding, user embedding, predicted reward and uncertainty and entropy estimation.

\textbf{Item Embedding}, $\mathbf{e}_{i}$, comes from the item ID and item features like the tags for music and categories for movies. 
\begin{equation}
\abovedisplayskip=2pt
\belowdisplayskip=2pt
    \mathbf{e}_{i} = f_I(i_{id}, \mathbf{F}^i_{id}), 
\end{equation}
where $f_I$ is the item encoder and $\mathbf{F^i}$ is the item features.

\textbf{User Embedding}, $\mathbf{e}_u$, is similar to item embedding for user ID and other static features. It is also considered as time-invariant.
\begin{equation}
\abovedisplayskip=2pt
\belowdisplayskip=2pt
\label{item_emb}
    \mathbf{e}_u = f_U(u_{id}, \mathbf{F}^u_{id}),
\end{equation}
where $f_U$ is the user encoder and $\mathbf{F}^u$ is the user features.

\textbf{Reward Prediction} is the core output of the world model. If we formulate the feedbacks of all the users towards all the items as a matrix, this matrix is usually highly sparse based on the offline interaction data, which is one typical characteristic of RS. The function of reward prediction, $\hat{r}$, is to answer the ``what if'' queries and complete the matrix. But the quality of this matrix complement and its influence have not been investigated. In recent methods~\cite{gao2023dorl,gao2023cirs}, the reward is the average of multiple world models
\begin{equation}
\abovedisplayskip=2pt
\belowdisplayskip=2pt
\label{r1}
    \bar{r} = \frac{1}{M} \sum^M_{j=1} W_j(\mathbf{e}_i, \mathbf{e}_u),
\end{equation}
where $M$ is the number of world models, and $W_j$ is the $j$-th one.
\textbf{Uncertainty Penalty} is widely used in offline RL to estimate the risk of taking a certain action. One representative direction utilizes an ensemble of world models to calculate the distances within the ensemble as a measure of uncertainty. In DORL, it formulates the world model as a Gaussian probabilistic model (GPM) and calculates the uncertainty of one interaction, $\mathbf{x}$, as $P_U(\mathbf{x}) := \max_{k \in \mathbf{E}} \sigma^2_{\theta_k}$, where $k$ is the index in world model ensemble $\mathbf{E}$, and $\sigma^2_{\theta_k}$ is the variance of corresponding GPM. Thus, the estimated reward in Equation (\ref{r1}) has been changed to:
\begin{equation}
\abovedisplayskip=2pt
\belowdisplayskip=2pt
\label{r2}
    \bar{r}' = \bar{r} - \lambda_U P_U,
\end{equation} where $\lambda_U$ is the uncertainty coefficient to adjust its scale and the uncertainty estimation naturally binds with the world model. If the world model is inaccurate, the uncertainty will be impacted. 

\textbf{Entropy Penalty} is a critical contribution of DORL. Though it is calculated with the offline data rather than based on the world model, it is also implemented in the first stage of training. Thus, we introduce it here to ensure a complete illustration and consistency with the implementation of DORL. The entropy penalty is calculated as the summation of a $k-$order entropy, which takes recent $k$ interactions as a pattern and counts the frequency of the next item matching the current pattern. 
\begin{equation}
\abovedisplayskip=2pt
\belowdisplayskip=2pt
\label{pe}
    P_E = -D_{\text{KL}}(\pi_{\beta}(\cdot|\mathbf{s})||\pi_u(\cdot|\mathbf{s})),
\end{equation}
where $\pi_{\beta}$ and $\pi_u$ are the behavior policy and the uniform distribution, respectively. It is an effective penalty that encourages the policy to explore the world model and improves the cumulative rewards. The final reward function during the policy learning is:
\begin{equation}
\abovedisplayskip=2pt
\belowdisplayskip=2pt
    r = \bar{r} - \lambda_U P_U + \lambda_E P_E,
\end{equation} 
where $\lambda_E$ is the coefficient for the entropy used to control its scale. In this way, DORL successfully alleviates the Matthew Effect.
\vspace{-0.5em}

\subsection{State Tracker}
\vspace{-0.2em}
State tracker models the transition function, $\mathbf{s'} = f(\mathbf{s}, \mathbf{a}, r)$, during the policy learning. It is also the state encoder that combines both the static and dynamic information as: 
\begin{equation}
\abovedisplayskip=2pt
\belowdisplayskip=2pt
    \mathbf{s'} = f(\mathbf{s}, \mathbf{e}_i, \mathbf{a}, \hat{r}|\mathbf{e}_u), 
\end{equation}where $\mathbf{u}_i$, $\hat{r}$, and $\mathbf{a}$ come from the world model introduced in the last section. In the implementation of DORL, the static user embedding is not considered in its state tracker. In addition, the $f$ is implemented as the average of recent $w$ item embedding concatenating with the estimated rewards $\hat{r}$, where $w$ is named as window size:
\begin{equation}
\abovedisplayskip=2pt
\belowdisplayskip=2pt
\label{state_repr}
    \mathbf{s_{t+1}} = \frac{1}{w} \sum^t_{j=t-w+1}[\mathbf{a_j} \oplus \hat{r}(\mathbf{s_j}, \mathbf{a_j})].
\end{equation}
Since the average tracker loses the order information, in our implementation, we turn to an attention tracker which enhances the cumulative reward in most cases. 
\vspace{-0.5em}

\subsection{Action Representation}
\vspace{-0.2em}
In the current setting, the item embedding from the world model is used to initialize action representation during policy learning since each item corresponds to one action. 
\begin{equation}
\abovedisplayskip=1pt
\belowdisplayskip=1pt
\label{act_repr}
    \mathbf{a_t} =  \mathbf{e}_{i}.
\end{equation}
On one hand, the action representation also influences policy learning. On the other hand, the inaccurate reward estimation from DeepFM intrigues us to doubt the quality of item embeddings. Thus, we find that replacing the current item embedding with a random initialization sampled from a standard normalization can enhance the cumulative rewards in many settings. Since this interesting direction is not our current focus, we leave it as our future direction.

\subsection{Policy Learning Pipeline}
\vspace{-0.2em}
\label{pipline}
In this part, we illustrate the policy learning process by interacting with the world model. Since the states are continuous, policy gradient or actor-critic algorithms are often used as baselines. We follow the implementation of DORL to build our method based on Advantage Actor-Critic~\cite{mnih2016a2c} (A2C).

The general pipeline for actor-critic algorithms is: The agent interacts with the world model to sample interaction trajectories, $\tau = \{(\mathbf{s_t}, \mathbf{a_t}, \mathbf{s_{t+1}}, r_t)\}$, using current policy $\pi_\theta$ where $\theta$ refers to the parameters representing the policy network. Then, the critic estimates the value function, either $V(\mathbf{s})$ or $Q(\mathbf{s}, \mathbf{a})$, for the current $\pi_\theta$. It uses $\tau$ to update its estimation, aiming to minimize the two parts between the Bellman equation (\eqref{v_s} or \eqref{q_sa}). Next, the actor updates $\theta$ by ascending the gradient of the expected cumulative reward estimated by the critic, aiming to maximize it. To sum up, the objective function of the critic is:
\begin{equation}
\abovedisplayskip=2pt
\belowdisplayskip=2pt
\label{critic}
    L_{\text{critic}} = \mathbb{E}_\tau\left[\left(r_t+\gamma \max _{\mathbf{a^{\prime}}} Q\left(\mathbf{s_{t+1}}, \mathbf{a^{\prime}} ; \phi\right)-Q\left(\mathbf{s_t}, \mathbf{a_t} ; \phi\right)\right)^2\right],
\end{equation} 
where $\phi$ represents the parameters of the critic network estimating $Q(s, a)$. The objective of the actor is:
\begin{equation}
\label{actor}
    J_{\text{actor}}=\mathbb{E}_\tau\left[\log \pi_\theta\left(\mathbf{a_t} \mid \mathbf{s_t}\right) \cdot A\left(\mathbf{s_t}, \mathbf{a_t}\right)\right].
\end{equation}
In A2C, the advantage function, $A_{\pi_\theta} = Q_{\pi_\theta}(\mathbf{s}, \mathbf{a}) - V_{\pi_\theta}(\mathbf{s})$, is introduced to accelerate the learning process.
\vspace{-0.5em}

\subsection{Reward Shaping}
\vspace{-0.2em}
The reward functions for most existing world models are inaccurate. The reason is also one of the main challenges of RS: the offline data is too sparse to comprehensively reflect users' true feedback. To overcome this problem, we need to dig out the intrinsic patterns in the offline data. Intuitive thought is that within a short time interval, some users may exhibit similar interests in a group of items, forming a cluster. Within this user cluster and the group of items, the feedback of a specific user toward a certain item may be inferred from the other users' feedback. Thus, based on our exploration and observation, a non-parametric reward shaping method that accounts for the specialties of RS is proposed to improve the prediction of the reward functions. 

Unlike training the reward model from extremely sparse offline data, user representation learning is comparably more informative thanks to the existence of static side information from users~\cite{wang2020aspect}. In addition, the user interaction history or the counterfactual interaction history can also be used to identify the users. We use this user information as their indicator features, denoted as $u$ for brevity. For each user in the evaluation environment, we utilize its indicator features to retrieve similar users in the training environment based on an appropriate distance metric using a clustering method. Here we adopt a soft-label \textit{kNN} to discover the user clusters in the offline data. Then, the reward correction is estimated by aggregating the feedback of these nearest neighbors: 
\begin{equation}
\abovedisplayskip=2pt
\belowdisplayskip=2pt
\label{knn}
    \tilde{r}_u(\mathbf{s},\mathbf{a}) = \frac{1}{k} \sum_{u' \in k\text{NN}(u)}r_{u'}(\mathbf{s},\mathbf{a}),
\end{equation}

In this paper, we use averaging as our aggregation function. The choice of $k$ depends on some statistics of the offline datasets. They will be elaborated in the experiment section.
\vspace{-0.5em}

\subsection{Uncertainty Penalty}
\vspace{-0.2em}
\label{sec:uncertainty}
As many current uncertainty estimations in offline RL for RecSys rely on an ensemble of world models~\cite{jeunen2021pessimistic,gao2023dorl}, the uncertainty penalty inevitably suffers from the inaccurate prediction of the reward functions. Thus, considering the special difficulties in our setting, we propose to penalize the distances between a user and its nearest neighbors in Eq.\eqref{knn} as the uncertainty. Then, we have, 
\begin{equation}
\abovedisplayskip=2pt
\belowdisplayskip=2pt
\label{p_u}
    \tilde{P}_U (u) = \frac{1}{k} \sum_{u' \in k\text{NN}(u)} d(u, u'),
\end{equation} 
where $d(\cdot)$ is the distance metric and we omit $u$ for brevity. Intuitively, it works as a complementary to our reward shaping method since it considers both the clustering quality and the conditions of the dataset to enhance the generalization. When the dataset is extremely sparse, the retrieved nearest neighbors for mutual inference may be less representative. Thus, our uncertainty penalty can effectively reduce the chance of making risky decisions. Specifically, we use the cosine distance as $d$ in our implementation. Extensive experiments about the design and estimation of uncertainty penalty are conducted in Section \ref{section: Pu}.
\vspace{-0.5em}

\subsection{Algorithm Overview}
\vspace{-0.2em}
Now the reward function for policy learning is derived as, 
\begin{equation}
\abovedisplayskip=2pt
\belowdisplayskip=2pt
\label{rew_fc}
    r = \tilde{r} \times (1- \tilde{P}_U) + \lambda_E P_E.
\end{equation}
\begin{algorithm}[!t]
\caption{Offline A2C Training with $k$NN Reward Shaping.}
\label{alg:Framwork}
\begin{algorithmic}[1] 
\REQUIRE ~~\\ 
    Offline dataset $D$; Learned world model $E$ (Environment); Total training epoch $K$; Number of trajectories in each epoch $N$
\ENSURE ~~\\ 
   The learned policy, $\pi_\theta$;
    \STATE Initialize the policy and critic network parameters, $\theta$ and $\phi$;
    \STATE Calculate the $k$NN-based reward $\tilde{r}$ by Eq. \eqref{knn}; 
    \STATE Calculate the $k$NN-based uncertainty $\tilde{P}_U$ by Eq. \eqref{p_u}; 
    \STATE Calculate the entropy penalty $P_E$ by Eq. \eqref{pe}; 
    \STATE Derive the reward function $r$ by Eq. \eqref{rew_fc};
    \FOR{each epoch $k = 0,1,2, ..., K$}
    \STATE n=0
    \WHILE{n < N}
    \STATE The actor samples a trajectory $\tau$ with current policy $\pi_{\theta}$ by interacting with the environment $E$;
    \STATE Calculate advantages $A(\mathbf{s_t},\mathbf{a_t})=Q(\mathbf{s_t},\mathbf{a_t})-V(\mathbf{s_t})$ and the cumulative reward for each time step in $\tau$;
    \STATE Update the critic, \textit{i.e.,} $\phi$, by Eq. \eqref{critic};
    \STATE Update the actor, \textit{i.e.,} $\theta$, by ascending the policy gradient of Eq. \eqref{actor};
    \STATE $n = n+1$;
    \ENDWHILE
    \ENDFOR
\RETURN the policy $\pi_\theta$; 
\end{algorithmic}
\end{algorithm}
\noindent We summarize the overall training process in the Algorithm~\ref{alg:Framwork}. In the evaluation process, given a user and its recent interaction trajectories, the state tracker calculates its state representation with Equation~\eqref{state_repr}. Then, this representation along with the action representation obtained through Equation~\eqref{act_repr} are fed into the learned policy $\pi_\theta$. The policy recommends one item to the user. This process lasts until the user quits interaction.
\vspace{-0.5em}

\section{Performance Lower Bound}
In the simulated MDP from the world model, $\hat{G} = <\mathbf{S}, \mathbf{A}, T, \hat{r},\\ \gamma>$, the $\hat{r}$ is estimated by our reward shaping method. Under the actor-critic framework, our analysis mainly focuses on the Bellman Error in the critic since ideally, if the critic can perfectly estimate the ground truth value function, the actor can make the best decisions from a finite action space accordingly. 

We are interested in the closeness of the estimated value function learned on $\hat{G}$ and the underlying ground truth value function corresponding to $G$ in the evaluation environment: $|V^*-V_{\pi}|, \forall s \in \mathbf{S}$. To conduct meaningful analysis, we need to make some mild assumptions on the distances between a user and its near neighbors. We denote $\hat{Q}^*$ as the optimal state-action value function of $\hat{G}$, $\mathcal{T}$ as the Bellman Operator, and $d$ as the distance metric for selecting nearest neighbors.
\vspace{-0.3em}

\begin{assumption}(Lipschitz continuity)
\label{assump}
    For any two transitions $(\mathbf{s}, \mathbf{a}), (\mathbf{s'}, \mathbf{a'}) \in S \times A$, the difference of their estimated value function after value iteration is Lipschitz continuous:
    \begin{equation}
    \abovedisplayskip=2pt
    \belowdisplayskip=2pt
        |\mathcal{T}\hat{Q}(\mathbf{s}, \mathbf{a}) - \mathcal{T}\hat{Q}(\mathbf{s'}, \mathbf{a'})| \leq L \cdot d[(\mathbf{s}, \mathbf{a}), (\mathbf{s'}, \mathbf{a'})].
    \end{equation}
\end{assumption}

where $L$ is the Lipschitz constant. This assumption bounds the differences in the value function estimation between any two neighborhood state-action pairs under the Bellman Operation of the evaluation environment. Based on the statistics of the offline data, we also need to consider the quality of the clustering. We use $d_m := \max_{(\mathbf{s_i},\mathbf{a_i})}d[(\mathbf{s_i},\mathbf{a_i}), 
$ $(\mathbf{s_j}, \mathbf{a_j})], (\mathbf{s_i},\mathbf{a_i}) \in \mathbf{S} \times \mathbf{A}, (\mathbf{s_j},\mathbf{a_j}) \in kNN((\mathbf{s_i}, \mathbf{a_i}))$ to represent the maximal distance between the feedback of a user and that of its nearest neighbors. In addition, the size of offline data and the number of neighbors are noted as $N$ and $K$, respectively. Then, we can derive the discrepancy between the estimated value function and the ground truth one in the following theorem. 
\vspace{-0.3em}

\begin{theorem}
\label{thm}
For offline data of size $N$, $\hat{Q}^*$ is its optimal value function with respect to its world model. If $\mathcal{T}\hat{Q}^*$ is $L-$smooth, then with probability at least 1 - $\delta$,
\begin{equation}
\abovedisplayskip=2pt
\belowdisplayskip=2pt
    |V_{\hat{\pi}^*} - V^*| \leq \frac{2\left(L \cdot d_m+ Q_m \cdot \epsilon(k, N, \delta)\right)}{1-\gamma},
\end{equation}
\end{theorem}
where $\epsilon$ is a little quantity and $Q_m$ is the maximal Q-value. This theorem provides our reward shaping method with a theoretical guarantee that when the users' behaviors are not too irrelevant to form a meaningful clustering and the sparsity of the offline data is limited, the policy learned with the world model is lower bounded.
\vspace{-0.5em}

\subsection{Proof Sketch}
\vspace{-0.2em}
In this part, we proof sketch of the Theorem ~\eqref{thm}. Before it starts, a Lemmas from ~\cite{pazis2013pac} will be introduced at first. 

\textbf{Lemma 1.} Let $\epsilon_{-} \geq 0$ and $\epsilon_{+} \geq 0$ be constants such that $\forall(s, a) \in(\mathbf{S}, \mathbf{A}),-\epsilon_{-} \leq Q(s, \bar{a})-\mathcal{T} Q(s, a) \leq \epsilon_{+}$. The return $V^\pi$ from the greedy policy over $Q$ satisfies:
\begin{equation}
\abovedisplayskip=2pt
\belowdisplayskip=2pt
\label{lemma}
    \forall s \in \mathcal{S}, V^\pi(s) \geq V^*(s)-\frac{\epsilon_{-}+\epsilon_{+}}{1-\gamma}
\end{equation}

\textbf{Proof:} We denote the value function for the real environment $G$ and the world model $\hat{G}$ as $Q$ and $\hat{Q}$. Applying the Bellman Operator on the same state-action pair of two environments is different as their reward functions differ. Thus, we use $\mathcal{T}$ and $\mathcal{\hat{T}}$ to denote them. To build a bridge between these two operators, we define a cross-environment Bellman Operator as a bridge to bind $G$ and $\hat{G}$:
\begin{equation}
\abovedisplayskip=2pt
\belowdisplayskip=2pt
    \mathcal{\tilde T}Q(s,a) = \hat{r} + \gamma \max_{a'}Q(s', a') + \mathcal{T}\hat{Q}(s, a) - \mathcal{T}\hat{Q}(s_i, a_i),
\end{equation} where $(s_i,a_i) \in k\text{NN}(s,a)$

Then, we decompose the Bellman Error as:
\begin{equation}
\abovedisplayskip=2pt
\belowdisplayskip=2pt
    \hat{Q}(s,a) - \mathcal{T}\hat{Q}(s,a) = \hat{Q}(s,a) - \mathcal{\tilde T}\hat{Q}(s, a) + \mathcal{\tilde T}\hat{Q}(s, a) - \mathcal{T}\hat{Q}(s,a).
\end{equation}

We focus on the right-hand side, taking the first two terms as one unit:
\begin{equation}
\abovedisplayskip=2pt
\belowdisplayskip=2pt
\mathcal{I} =\frac{1}{k} \sum_{i \in k N N(s, a)} [\mathcal{T}\hat{Q}(s, a) - \mathcal{T}\hat{Q}(s_i, a_i)];
\end{equation}

Due to our assumption~\ref{assump} about the Lipschitz continuity, we have:
\begin{equation}
\abovedisplayskip=2pt
\belowdisplayskip=2pt
    -L \cdot d_m \leq \mathcal{I} \leq L \cdot d_m;
\end{equation}

Now we need to bound $\mathcal{II} = \mathcal{\tilde T}\hat{Q}(s, a) - \mathcal{T}\hat{Q}(s,a)$. We have
\begin{equation}
\abovedisplayskip=2pt
\belowdisplayskip=2pt
E[\mathcal{\tilde{T}} \hat{Q}(s, a)] =\mathcal{T} \hat{Q}(s, a) 
\end{equation}

We denote $C(N, K)$ as the largest number of clusters in a dataset. According to the Hoeffding inequality, given real number $\delta \in (0,1)$, we have,
\begin{equation}
\abovedisplayskip=2pt
\belowdisplayskip=2pt
    P\left(|\mathcal{T} \hat{Q}(s, a)-\mathcal{T} \hat{Q}(s, a)| \geqslant Q_m \sqrt{\frac{1}{2 k} \ln \frac{2C(N, k)}{\delta}}\right) \leq \delta,
\end{equation} 
where $Q_m$ is the maximal Q-value.

Then, we obtain that with probability at least $1-\delta$, $|\mathcal{II}| \leq \epsilon (k, N, \delta)$. Putting $\mathcal{I}$ and $\mathcal{II}$ together, we have,
\begin{equation}
\abovedisplayskip=2pt
\belowdisplayskip=2pt
    -(L \cdot d_m + \epsilon (k, N, \delta)) \leq \hat{Q}(s, a) - \mathcal{T}\hat{Q}(s,a) \leq L \cdot d_m + \epsilon (k, N, \delta).
\end{equation}

Use Eq. \eqref{lemma} in Lemma 1, the Theorem is proved.
\vspace{-0.5em}

\section{Experiment}
In this section, we begin with introducing the experiment setup including the datasets, evaluation metric, the state-of-the-art method, and other baselines. Then, we investigate the following questions:

\noindent$\bullet$ (RQ1) How does ROLeR perform compared with other baselines?

\noindent$\bullet$ (RQ2) How do the reward shaping and uncertainty penalty in ROLeR contribute to its performance?

\noindent$\bullet$ (RQ3) What is the most effective design of the uncertainty penalty?

\noindent$\bullet$ (RQ4) What is the impact of the world model in ROLeR?

\noindent$\bullet$ (RQ5) Is ROLeR robust to the critical hyperparameters? 
\vspace{-0.5em}

\subsection{Setup}
\vspace{-0.2em}
\subsubsection{Datasets}
We conduct experiments on four datasets including two newly proposed challenging short-video interaction datasets and two typical datasets. The statistics are listed in Table \ref{tb:dataset}.

\textbf{KuaiRec}~\cite{gao2022kuairec} is a video dataset that contains a fully-observed user-item interaction matrix used for evaluation. Within this matrix, each user's feedback towards any item is known. The normalized viewing time of a video is used as a reward signal. The training data is from the standard interaction history containing popularity bias, which makes it difficult to learn a recommendation policy.

\textbf{KuaiRand-Pure}~\cite{gao2022kuairand} is also a video dataset that contains user-item interaction collected by randomly inserting the target videos in the standard recommendation stream. This part is used to simulate a fully observed interaction matrix like KuaiRec for evaluation, while the data collected in the standard stream serves as the training set. The \textit{"is\_click"} signal is used as the reward. 

\textbf{Coat}~\cite{schnabel2016coat} is a shopping dataset. It consists of the user ratings, a five-point scale, on the self-selected items and uniformly sampled ones. These two parts of data are used for training and testing, respectively. The ratings are used as reward signals.

\textbf{Yahoo}~\cite{marlin2009yahoo} is a music dataset with training set and testing set collection similar to Coat. The ratings on the user-selected items are used for training, while those on the randomly picked items are for testing. Similarly, the ratings are treated as reward signals.
\vspace{-0.3em}
\begin{table}
    \centering
    \caption{Dataset Statistics.}
    \vspace{-1.1em}
    \label{tb:dataset}
    {
    \renewcommand{\arraystretch}{0.3}
    \small
    \begin{tabular}{ccccc}
    \toprule
    Dataset                   & Usage & No. of Users & No. of Items & \makecell{Density}   \\
    \midrule
    \multirow{2}{*}{KuaiRec}  & Train & 7176         & 10728        & 16.277\%             \\
    \cmidrule{2-5}
                              & Test  & 1411         & 3327         & 99.620\%             \\
    \midrule
    \multirow{2}{*}{KuaiRand} & Train & 27285        & 7551         & 0.697\%              \\
    \cmidrule{2-5}
                              & Test  & 27285        & 7583         & 0.573\%             \\
    \midrule
    \multirow{2}{*}{Coat}     & Train & 290          & 300
       & 8.046\%              \\
    \cmidrule{2-5}
                              & Test & 290           & 300
       & 5.287\%              \\
    \midrule
    \multirow{2}{*}{Yahoo}  & Train & 15400        & 1000
       & 2.024\%              \\
    \cmidrule{2-5}
                              & Test & 5400          & 1000
       & 1.000\%              \\
    \bottomrule
    \end{tabular}
    }
\end{table}

\subsubsection{Evaluation}
In the scope of RL for recommender systems, the objective of RS is to maximize long-term user satisfaction, which can be formulated as the return, \textit{i.e.,} cumulative reward. For the detailed evaluation setting, we follow the same manners as in DORL~\cite{gao2023dorl} and~\cite{gao2023cirs}: The same item will not be recommended twice to a user. A quitting mechanism is applied as the termination condition: the interaction will terminate when a user receives $M$ items of the same category in the recent $N$ transitions. We keep $M=0, N=4$ as in DORL, which means for every four transitions, the items' categories should differ. The maximum user-item interaction length is set to 30 and the cumulative reward is evaluated after every training epoch on 100 testing trajectories. The final cumulative reward is averaged across 200 epochs. In addition, to enable in-depth analysis, the cumulative reward is decomposed into the interaction length and single-step recommendation reward. They are regarded as metrics as well. Majority category domination (MCD) was originally proposed in DORL as a reference of the Matthew effect, which should be controlled in a suitable range. For readability, we still overbold the smallest MCD in each dataset. In addition, due to the definition of \textit{most popular categories} in DORL, the calculation of MCD is not supported on Coat and Yahoo.

While the testing set of KuaiRec is fully observable and can be directly used as the reward function during policy evaluation, the reward functions for the other three datasets are estimated with the testing data by DeepFM models. To ensure a fair comparison, we use exactly the same reward functions as in DORL on all datasets.

All experiments are conducted an NVIDIA RTX™ A6000 with 48 GB GDDR6. Each of these requires less than 3 GPU hours, while each trial of ROLeR requires no more than 2 GPU hours.
\begin{table*}[!t]
\caption{The performance summary of all methods on KuaiRand and KuaiRec. (Bold: best; Underline: runner-up)}
\vspace{-1.1em}
\label{tb:main_result1}
\centering
\resizebox{\linewidth}{!}{
\renewcommand\arraystretch{0.85}
\begin{tabular}{c|c|c|c|c|c|c|c|c}
\toprule
\multirow{2}{*}{Exp} & \multicolumn{4}{c|}{KuaiRand} & \multicolumn{4}{c}{KuaiRec} \\
\cmidrule{2-5}
\cmidrule{6-9}
 & $\text{R}_\text{tra}$ $\uparrow$ & $\text{R}_\text{each}$ $\uparrow$ & Length $\uparrow$ & MCD & $\text{R}_\text{tra}$ $\uparrow$ & $\text{R}_\text{each}$ $\uparrow$ & Length $\uparrow$ & MCD \\
\midrule
\midrule
UCB & $1.6510\pm 0.1515$ & $0.3725\pm 0.0278$ & $4.4312\pm 0.2121$ & $0.7886\pm 0.0235$ & $3.6059\pm 0.6092$ & $0.8531\pm 0.1145$ & $4.2190\pm 0.3892$ & $0.8112\pm 0.0582$ \\
$\epsilon$-greedy & $1.7109\pm 0.1258$ & $0.3510\pm 0.0251$ & $4.8804\pm 0.2700$ & $0.7735\pm 0.0239$ & $3.5152\pm 0.7315$ & $0.8276\pm 0.1286$ & $4.2186\pm 0.4049$ & $0.8226\pm 0.0482$ \\
SQN & $0.9117\pm 0.9292$ & $0.1818\pm 0.0584$ & $4.6007\pm 3.7125$ & $0.6208\pm 0.1865$ & $4.6730\pm 1.2149$ & \underline{$0.9125\pm 0.0551$} & $5.1109\pm 1.2881$ & $0.6860\pm 0.0931$ \\
CRR & $1.4812\pm 0.1236$ & $0.2258\pm 0.0151$ & $6.5613\pm 0.3519$ & $0.7326\pm 0.0187$ & $4.1631\pm 0.2535$ & $0.8945\pm 0.0365$ & $4.6541\pm 0.2150$ & $0.8648\pm 0.0168$ \\
CQL & $2.0323\pm 0.1070$ & $0.2258\pm 0.0119$ & $9.0000\pm 0.0000$ & $0.7778\pm 0.0000$ & $2.5062\pm 1.7665$ & $0.6843\pm 0.2279$ & $3.2239\pm 1.3647$ & $0.3858\pm 0.3853$ \\
BCQ & $0.8515\pm 0.0523$ & $0.4246\pm 0.0164$ & $2.0050\pm 0.0707$ & $0.9983\pm 0.0236$ & $2.1234\pm 0.0815$ & $0.7078\pm 0.0272$ & $3.0000\pm 0.0000$ & $0.6667\pm 0.0000$ \\
MBPO & $10.9325\pm 0.9457$ & $0.4307\pm 0.0210$ & $25.3446\pm 1.8190$ & $0.3061\pm 0.0403$ & $12.0426\pm 1.3115$ & $0.7701\pm 0.0290$ & $15.6461\pm 1.6373$ & \underline{$0.3621\pm 0.0465$} \\
IPS & $3.6287\pm 0.6763$ & $0.2163\pm 0.0141$ & $16.8213\pm 3.1824$ & $\mathbf{0.2010\pm 0.1156}$ & $12.8326\pm 1.3531$ & $0.7673\pm 0.0234$ & $16.7270\pm 1.6834$ & $\mathbf{0.2150\pm 0.0644}$ \\
MOPO & $10.9344\pm 0.9634$ & \underline{$0.4367\pm 0.0193$} & $25.0019\pm 1.8911$ & $0.3433\pm 0.0289$ & $11.4269\pm 1.7500$ & $0.8917\pm 0.0505$ & $12.8086\pm 1.8502$ & $0.4793\pm 0.0619$ \\
DORL & \underline{$11.8500\pm 1.0361$} & $0.4284\pm 0.0223$ & \underline{$27.6091\pm 2.1208$} & \underline{$0.2960\pm 0.0356$} & \underline{$20.4942\pm 2.6707$} & $0.7673\pm 0.0264$ & \underline{$26.7117\pm 3.4190$} & $0.3792\pm 0.0149$ \\
\midrule
ROLeR & $\mathbf{13.4553\pm 1.5086}$ & $\mathbf{0.4574\pm 0.0332}$ & $\mathbf{29.2700\pm 2.3225}$ & $0.4049\pm 0.0356$ & $\mathbf{33.2457\pm 2.6403}$ & $\mathbf{1.2293\pm 0.0511}$ & $\mathbf{27.0131\pm 1.3986}$ & $0.4439\pm 0.0212$ \\
\midrule
\midrule
\textcolor{gray}{GT Reward} & \textcolor{gray}{$14.3689\pm 1.9708$} & \textcolor{gray}{$0.4993\pm 0.0488$} & \textcolor{gray}{$28.5582\pm 2.4114$} & \textcolor{gray}{$0.4109\pm 0.0397$} & \textcolor{gray}{$36.7475\pm 3.4738$} & \textcolor{gray}{$1.5600\pm 0.0405$} & \textcolor{gray}{$23.5653\pm 2.1824$} & \textcolor{gray}{$0.5594\pm 0.0267$} \\
\bottomrule
\end{tabular}
}
\end{table*}
\begin{table*}[!t]
\vspace{-0.3em}
\caption{The performance summary of all methods on Coat and Yahoo. (Bold: best; Underline: runner-up). Due to the definition of MCD, its calculation is not supported on Coat and Yahoo.}
\vspace{-1.1em}
\label{tb:main_result2}
\centering
\resizebox{0.78\linewidth}{!}{
\renewcommand\arraystretch{0.85}
\begin{tabular}{c|c|c|c|c|c|c}
\toprule
\multirow{2}{*}{Exp} & \multicolumn{3}{c|}{Coat} & \multicolumn{3}{c}{Yahoo} \\
\cmidrule{2-4}
\cmidrule{5-7}
 & $\text{R}_\text{tra}$ $\uparrow$ & $\text{R}_\text{each}$ $\uparrow$ & Length $\uparrow$ & $\text{R}_\text{tra}$ $\uparrow$ & $\text{R}_\text{each}$ $\uparrow$ & Length $\uparrow$ \\
\midrule
\midrule
UCB & $73.6713\pm 1.8105$ & $2.4557\pm 0.0604$ & $30.0000\pm 0.0000$ & \underline{$66.7578\pm 1.2539$} & \underline{$2.2253\pm 0.0418$} & $30.0000\pm 0.0000$ \\
$\epsilon$-greedy & $72.0042\pm 1.6054$ & $2.4001\pm 0.0535$ & $30.0000\pm 0.0000$ & $64.3439\pm 1.2911$ & $2.1448\pm 0.0430$ & $30.0000\pm 0.0000$ \\
SQN & $72.6142\pm 2.0690$ & $2.4205\pm 0.0690$ & $30.0000\pm 0.0000$ & $57.7270\pm 5.7506$ & $1.9242\pm 0.1917$ & $30.0000\pm 0.0000$ \\
CRR & $67.3830\pm 1.6274$ & $2.2461\pm 0.0542$ & $30.0000\pm 0.0000$ & $57.9941\pm 1.6752$ & $1.9331\pm 0.0558$ & $30.0000\pm 0.0000$ \\
CQL & $68.9835\pm 1.8659$ & $2.2995\pm 0.0622$ & $30.0000\pm 0.0000$ & $62.2909\pm 3.3466$ & $2.0764\pm 0.1116$ & $30.0000\pm 0.0000$ \\
BCQ & $68.8012\pm 1.7627$ & $2.2934\pm 0.0588$ & $30.0000\pm 0.0000$ & $61.7388\pm 1.7808$ & $2.0580\pm 0.0594$ & $30.0000\pm 0.0000$ \\
MBPO & $71.1930\pm 2.0943$ & $2.3731\pm 0.0698$ & $30.0000\pm 0.0000$ & $64.5500\pm 2.1567$ & $2.1517\pm 0.0719$ & $30.0000\pm 0.0000$ \\
IPS & \underline{$73.8872\pm 1.8417$} & \underline{$2.4629\pm 0.0614$} & $30.0000\pm 0.0000$ & $57.8499\pm 1.7955$ & $1.9283\pm 0.0599$ & $30.0000\pm 0.0000$ \\
MOPO & $71.1805\pm 2.0560$ & $2.3727\pm 0.0685$ & $30.0000\pm 0.0000$ & $65.5098\pm 2.0996$ & $2.1837\pm 0.0700$ & $30.0000\pm 0.0000$ \\
DORL & $71.3992\pm 2.0640$ & $2.3800\pm 0.0688$ & $30.0000\pm 0.0000$ & $66.3509\pm 2.2237$ & $2.2117\pm 0.0741$ & $30.0000\pm 0.0000$ \\
\midrule
ROLeR & $\mathbf{76.1603\pm 2.1200}$ & $\mathbf{2.5387\pm 0.0707}$ & $30.0000\pm 0.0000$ & $\mathbf{68.3637\pm 1.8550}$ & $\mathbf{2.2788\pm 0.0618}$ & $30.0000\pm 0.0000$ \\
\midrule
\midrule
\textcolor{gray}{GT Reward} & \textcolor{gray}{$80.0895\pm 2.4545$} & \textcolor{gray}{$2.6696\pm 0.0818$} & \textcolor{gray}{$30.0000\pm 0.0000$} & \textcolor{gray}{$68.8791\pm 3.2867$} & \textcolor{gray}{$2.2960\pm 0.1096$} & \textcolor{gray}{$30.0000\pm 0.0000$} \\
\bottomrule
\end{tabular}
}
\vspace{-0.15em}
\end{table*}

\vspace{-0.8em}
\subsubsection{Baselines}
The baselines include five model-based offline RL methods, including two state-of-the-art, four model-free offline RL methods, and two vanilla bandit-based methods. To ensure a fair comparison, on each dataset, we use the same DeepFM~\cite{deepfm} model to learn the world model for model-based offline RL methods.

\noindent $\bullet$ \textbf{$\epsilon-$greedy}: This method uses the world model predicted reward as its reward function to make a decision with probability $1-\epsilon$ and takes actions randomly with probability $\epsilon$.

\noindent $\bullet$ \textbf{UCB}: Upper Confidence Bound bandit algorithm~\cite{lai1985ucb}, estimates an upper confidence bound and favors the actions with the highest confidence bounds to balance the exploration and exploitation.

\noindent $\bullet$ \textbf{SQN}: Self-supervised Q-learning~\cite{xin2020sqn}, adopts a dual-headed architecture where one focuses on the cross-entropy loss and the other is designed for RL tasks which we use for recommendation.

\noindent $\bullet$ \textbf{BCQ}: Batch-Constrained deep Q-learning~\cite{fujimoto2019bcq}, selectively updating policies based on high-confidence data, effectively sidelining ambiguous or uncertain data.

\noindent $\bullet$ \textbf{CQL}: Conservative Q-Learning~\cite{kumar2020cql}, constrains the usage of OOD data to update the state-action value function. 

\noindent $\bullet$ \textbf{CRR}: Critic Regularized Regression~\cite{wang2020crr}, updates the policy conditioning on the discrepancy in the behavior policies.

\noindent $\bullet$ \textbf{MBPO}: Model-Based Policy Optimization~\cite{janner2019mbpo}, is a baseline that trains an actor-critic policy on the learned world model.

\noindent $\bullet$ \textbf{MOPO}: Model-based Offline Policy Optimization~\cite{yu2020mopo}, proposes to penalize the uncertainty which is calculated through an ensemble of world models.

\noindent $\bullet$ \textbf{IPS}: Inverse Propensity Scoring~\cite{swaminathan2015ips}, is a statistics-based sample re-weighting method. An actor-critic policy is learned based on the world model after re-weighting the offline data.

\noindent $\bullet$ \textbf{DORL}: Debiased model-based Offline RL~\cite{gao2023dorl}, designs the uncertainty penalty and introduces the entropy penalty to encourage diverse recommendation and alleviate the Matthew Effect. Current state-of-the-art (SOTA) on both KuaiRec and KuaiRand.

\noindent $\bullet$ \textbf{CIRS}: Counterfactual Interactive Recommender System~\cite{gao2023cirs}, introduces a causal model to infer the user preference in the training of a PPO~\cite{schulman2017ppo}. CIRS is also a SOTA method on KuaiRec. The official repository of CIRS only covers KuaiRec in the four datasets.

\noindent $\bullet$ \textbf{GT Reward}: it trains the recommendation policy with the reward function from the testing environment, serving as a reference policy of one performance upper bound. The RL algorithm utilized is \textit{A2C}, which is the same as DORL and ROLeR.
\vspace{-0.8em}
\subsection{Overall Performance (RQ1)}
For our ROLeR, we have two key hyperparameters: the coefficients of the entropy penalty and the number of nearest neighbors. We simply tune $k$ from $5$ to $50$, $25$ to $200$, $10$ to $35$, and $10$ to $100$ on KuaiRec, KuaiRand, Coat, and Yahoo, respectively.

\begin{figure}[!t]
\centering
\includegraphics[trim=0cm 0cm 0cm 0cm, clip, width=0.97\columnwidth]{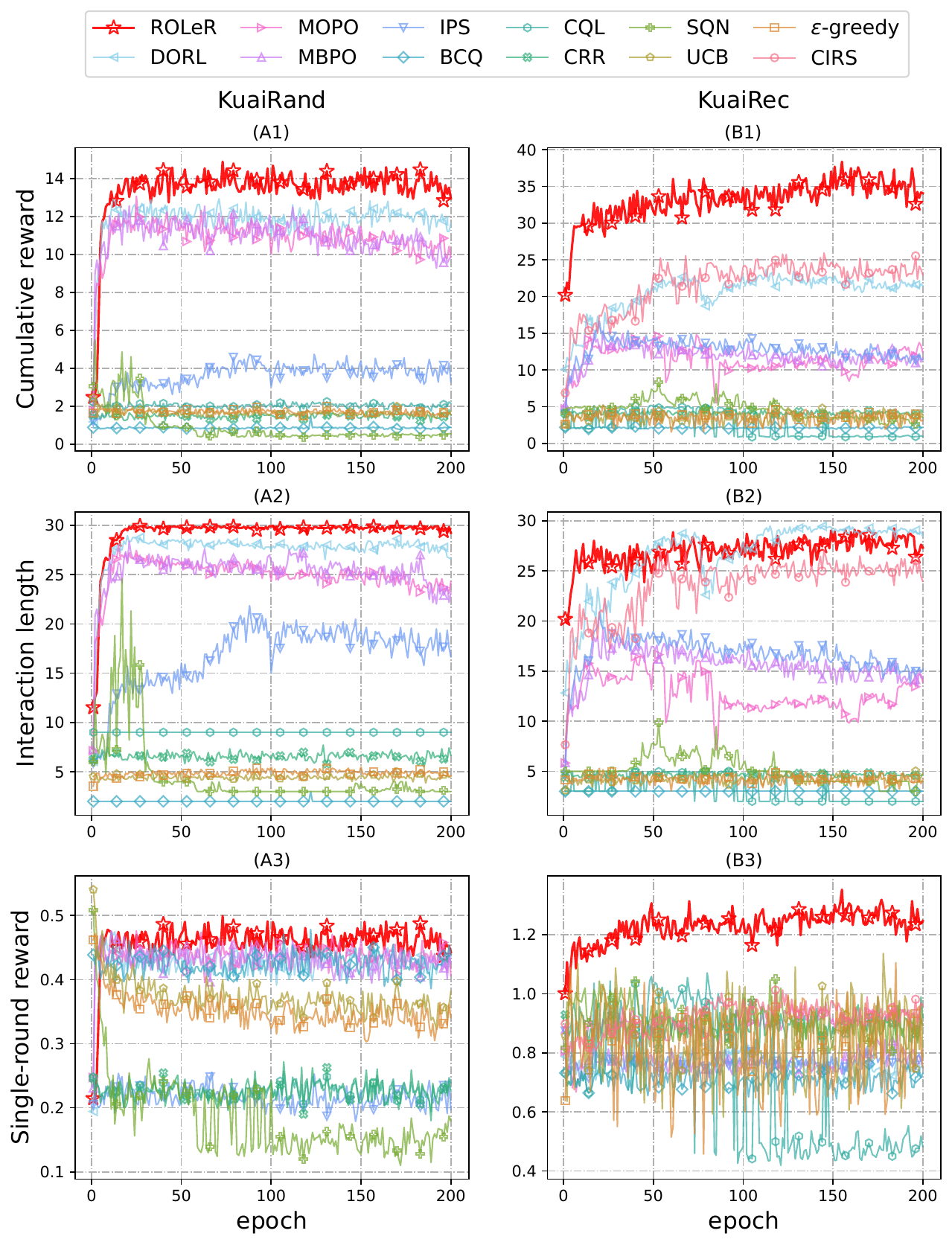}
\vspace{-1.7em}
\caption{The overall performance on KuaiRand and KuaiRec.}
\label{fig:main_result}
\end{figure}

The detailed performance on four datasets is listed in Table \ref{tb:main_result1} and \ref{tb:main_result2}. The corresponding curves for KuaiRand and KuaiRec are in Figure \ref{fig:main_result}. The first row is the main evaluation metric: cumulative reward followed by average interaction length and single-step reward as the second and third rows, respectively. We first analyze the results in a more specific view, \textit{i.e.,} the interaction length and the single-step reward. Then, we compile our observations on cumulative rewards, detailing the insights gleaned from the experiments.

ROLeR achieves the longest interaction on four datasets. In the second row of Figure \ref{fig:main_result}, it shows the interaction length of all methods on KuaiRand and KuaiRec. In KuaiRec, the curves in \textit{B3} can be classified into three levels. DORL, CIRS, and ROLeR are at the top level, converging to the interaction upper bound as 30. It is obvious that ROLeR can reach an interaction length of 20 at the first epoch, longer than other baselines. It showcases that ROLeR can avoid recommending risky items at an early training stage. The second level consists of other model-based offline RL methods, such as MBPO and IPS, while the model-free offline methods are difficult to reach at a length of 5. This observation is instructive in RecSys while both the model-based and model-free methods adopt a conservative design, the regularization on the discrepancy of the behavior policies and learning policy limits the actual action space, rendering the early quitting problem. However, learning a world model as a solution to this problem is not as effective as expected since most of them still struggle to reach the upper bound. Similar observations can be found on the KuaiRand dataset, but its performance is more differentiable. Combining the detailed values in \textit{Length} columns of Table \ref{tb:main_result1}, we believe that ROLeR can reach the longest interaction length is due to ROLeR's more precise understanding of the environment compared to other baseline methods. Specifically, the ROLeR policy is more effective at making recommendations when choosing between a risky high-reward item and a conservative low-reward item to ensure the interaction length.

As for Coat and Yahoo, due to the lesser overlap of item categories compared to those in KuaiRec and KuaiRand, all baseline methods can easily reach the maximum interaction length without triggering the termination condition as shown in the \textit{Length} columns in Table \ref{tb:main_result2}. So, we turn to the analysis of single-step reward.

ROLeR achieves the closest single-step reward to the reference policy trained by ground truth reward functions on all datasets.
The single-step reward range for KuaiRand is $0\sim1$ and $0\sim5$ for the other three datasets. Looking into the last row of Figure \ref{fig:main_result}, the ROLeR significantly outperforms all baseline methods by a large margin on KuaiRec. This is the major contribution of more accurate reward functions. It is also evidenced by the comparison in Figure \ref{fig:main_result_cossim}. 
On KuaiRand, Coat, and Yahoo, though the differences of single-step reward, shown by the $R_{each}$ columns in Table \ref{tb:main_result1} and \ref{tb:main_result2}, across all baselines are not so significant as KuaiRec, ROLeR can still utilize the "inconspicuous" differences to steadily accumulated to significant advantages in the long-term reward.

As for MCD, the experiment results show that it should be controlled to be less than $0.5$. In detail, a large MCD ($>0.5$) leads to shorter interaction trajectories, such as CQL and BCQ, resulting in limited cumulative rewards. On the other hand, the baselines with minimal MCD do not achieve the longest trajectories and highest cumulative reward, such as IPS, on both datasets. 

To summarize the superiority of ROLeR and draw valuable insights from the experiments, we focus on the cumulative reward in the first row of Figure ~\ref{fig:main_result} and the $R_{tra}$ columns in Table ~\ref{tb:main_result1} and ~\ref{tb:main_result2}. On four datasets, ROLeR significantly outperforms the baseline methods, yielding a sub-optimal performance to the reference policy. Comparing the performance of ROLeR with other baselines, the influence of the accuracy of the world models is vital. It verifies the motivation of this work. In addition, in the RL algorithm design, ROLeR uses the plain A2C while baseline methods such as CIRS and CQL use advanced algorithms like PPO~\cite{schulman2017ppo} and SAC\cite{haarnoja2018sac}. The result comparison demonstrates that before diving into a delicate policy design, pursuing an accurate world model for model-based methods in recommender systems is more fundamental and effective. 
\vspace{-0.8em}

\subsection{Ablation Study (RQ2)}
\vspace{-0.2em}
In this part, we investigate the significance of the proposed components: the non-parametric reward shaping and the uncertainty penalty on four datasets. Our experiments on KuaiRec are based on DORL with Transformer~\cite{vaswani2017attention} state tracker inspired by SASRec~\cite{kang2018sasrec} and Gaussian initialized item embedding, which are effective designs that can improve the performance of DORL. As for the other three datasets, these techniques cannot consistently increase the cumulative reward, so we continue to use the original DORL. Intuitively, ROLeR has three variants. One adapts the reward shaping but keeps the world model based uncertainty penalty. We denote this version as \textit{ROLeR without kr}. Another one keeps the world model predicted reward and changes to $k$NN-based uncertainty penalty denoted as \textit{ROLeR without ku}. As described in Section \ref{sec:uncertainty}, the uncertainty penalty collaborates with the $k$NN reward shaping and cannot be directly applied to the reward functions from world models. Thus, an alternative version, \textit{i.e.,} $\bar{r}-\lambda_U \tilde{P_U} + \lambda_E P_E$, is tested.
\begin{table}[]
\centering
\caption{Ablation studies on ROLeR's components with $R_{tra}$.}
\vspace{-0.8em}
\label{tab:ablation}
\scriptsize
\setlength{\tabcolsep}{2.7pt}
\renewcommand\arraystretch{0.7}
\begin{tabular}{l|c|c|c|c|c|c}
  \toprule
  Methods & ku & kr & KuaiRec & KuaiRand & Coat & Yahoo \\
  \midrule
  \midrule
  DORL & $\times$ & $\times$ & 24.4308 ± 1.6768 & 11.8500 ± 1.0335 & 71.3992 ± 2.0588 & 66.3509 ± 2.2181 \\
  \midrule
  \multirow{3}{*}{ROLeR} & $\surd$ & $\times$ & 23.4512 ± 1.4973 & 12.5136 ± 1.3781 & 72.6059 ± 1.8019 & 66.1680 ± 1.7706 \\
   & $\times$ & $\surd$ & 30.7430 ± 2.1178 & 12.7339 ± 1.3681 & 72.6055 ± 1.8186 & 68.0083 ± 1.8920 \\
   & $\surd$ & $\surd$ & 33.2457 ± 2.6337 & 13.4553 ± 1.5049 & 76.1603 ± 2.1147 & 68.3637 ± 1.8504 \\
  \bottomrule
\end{tabular}%
\end{table}
From Table \ref{tab:ablation}, either the reward shaping method or uncertainty penalty can improve the cumulative rewards compared to DORL. This observation is especially obvious for \textit{ROLeR without ku}, emphasizing on the contribution of the non-parametric reward shaping. Combining the results of the last two rows of this table, it is easy to find that the uncertainty penalty serves as a constructive complementary to the reward shaping, enabling ROLeR to suit different datasets. The superiority of ROLeR over the other variant demonstrates the effectiveness of its core components. 
\vspace{-0.8em}

\subsection{The Design of Uncertainty Penalty (RQ3)}
\vspace{-0.2em}
\label{section: Pu}
As illustrated in Section \ref{sec:uncertainty}, the uncertainty penalty can adjust the way and degree of reward rectification to influence generalization. To investigate the most effective design of the uncertainty penalty, we conduct extensive experiments about the possible variants of the current uncertainty penalty across four datasets, and the results are shown in Table \ref{tab:Pu}. We denote the distances between a user and its nearest neighbours as $d$ in this table, and $d_{min}/d_{avg}/d_{max}$ represents the minimum/mean/maximum of the distances. $\mathcal{N}(\cdot,\cdot)$ is the Gaussian distribution. In addition, the $\lambda$ for each variant is tuned and the optimal choice is different. We omit the subscript of $\lambda$ and the entropy penalty, $P_E$, for simplicity. The $r$ for all variants in this table denotes the reward function after our reward shaping.
\begin{table}[]
\centering
\caption{Uncertainty penalty design of ROLeR with $R_{tra}$.}
\vspace{-1.1em}
\label{tab:Pu}
\scriptsize
\setlength{\tabcolsep}{2.7pt}
\renewcommand\arraystretch{1}
\begin{tabular}{c|c|c|c|c}
  \toprule
  Methods & KuaiRec & KuaiRand & Coat & Yahoo \\
  \midrule\midrule
  DORL & 24.4308 ± 1.6768 & 11.8500 ± 1.0335 & 71.3992 ± 2.0588 & 66.3509 ± 2.2181 \\
  \midrule
  $r \times \frac{\lambda}{d}$ & 17.5296 ± 1.4775 & 5.9727 ± 0.7820 & 66.8383 ± 2.1119 & 52.7376 ± 1.7133 \\
   $\mathcal{N}(r, \lambda d)$ & 31.3588 ± 2.2046 & 11.5336 ± 1.2343 & 74.3904 ± 1.7987 & 66.8826 ± 1.7944 \\
   $r-\lambda d_{min}$ & 30.7615 ± 2.2389 & 12.6348 ± 1.3889 & 75.3494 ± 1.9297 & 67.2455 ± 1.8898 \\
   $r-\lambda d_{avg}$ & 31.5732 ± 2.1560 & 10.6609 ± 2.7965 & 75.1499 ± 2.0620 & 66.1912 ± 1.7949 \\
   $r-\lambda d_{max}$ & 31.4333 ± 2.0677 & 12.6874 ± 1.3913 & 72.3764 ± 1.8067 & 67.3516 ± 1.6982 \\
   $r-\lambda d$ & 33.2457 ± 2.6337 & 13.3310 ± 1.5288 & 74.7558 ± 1.8794 & 66.9811 ± 1.8107 \\
   \midrule
   $r \times (1-d)$ & 31.1419 ± 2.4650 & 13.4553 ± 1.5049 & 76.1603 ± 2.1147 & 68.3637 ± 1.8504 \\
  \bottomrule
\end{tabular}%
\end{table}
In the second line of Table \ref{tab:Pu}, we use the inverse of the distances as the uncertainty penalty to discount the probabilities of recommending highly uncertain items, \textit{i.e.,} $r \times \frac{\lambda}{d}$. As it tends to penalize the uncertainty too sharply, it yields the smallest cumulative reward across four datasets. In the next line, sampling from a Gaussian distribution with the predicted $r$ as mean and uncertainty penalty as the variance \textit{i.e.}, $\mathcal{N}(r, \lambda d)$, is also a conservative way to avoid risky recommendation. This variant implicitly penalizes the 
uncertainty by changing the deterministic reward into a Gaussian-distributional reward. Unfortunately, this variant does not perform robustly across four datasets. The following three variants, \textit{i.e.,} $r-\lambda d_{min}/d_{avg}/d_{max}$, serve as intuitive ways to adjust the degree of the uncertainty penalty. The corresponding results demonstrate that such a direct design is not sufficient to estimate the uncertainty in offline model-based RL for RecSys. The last second variant achieves the highest cumulative reward on KuaiRec. We believe it benefits from the special coverage of KuaiRec compared to the other datasets. As the training set of KuaiRec is comparably dense, the distances between a user and its nearest neighbors are capable of estimating the uncertainty. On the other three datasets, the last variant offers the highest return. Since we use the cosine distance as $d$, then $1-d$ indicates the cosine similarity. Thus, this variant uses the weighted average of nearest neighbors to refine the reward functions during mutual inference. To summarize, the uncertainty penalty design of ROLeR demonstrates robust performance across four datasets, suggesting its effectiveness compared to other variants. Considering the diverse properties of each dataset, how to automatically decide the most suitable uncertainty design will be a challenging yet meaningful topic that will be our future work.
\vspace{-0.8em}

\subsection{The Impact of World Model (RQ4)}
\vspace{-0.2em}
When summarizing the utility of the world model in policy learning, it trains the item embedding used as the initialization of action representation, predicts the user-item feedback as the estimated rewards, and estimates the uncertainty. In implementing ROLeR on KuaiRec, our reward model and uncertainty penalty are estimated from the offline data. We find that both initializing the action representation from a standard distribution and applying a Transformer state tracker can improve the cumulative reward in most of our testing settings as exemplified in Table \ref{tab:world_model}. In this table, \textit{att} and \textit{avg} mean the transformer state tracker and average state tracker, respectively. $a_G$ and $a_t$ mean whether to initialize the action representation at the beginning of policy training or not. We believe the inaccuracy of the world model does not only hurt the estimation of the reward function but the action representation. In addition, as recent interaction history is available, the former Avg. state tracker used in DORL can hardly capture the order information. That is the reason we turn to the Transformer state tracker on KuaiRec. However, this is not always the case for the other three datasets. The original testing matrix for KuaiRand is extremely sparse (see Table\ref{tb:dataset}). To make it a fully observed evaluation environment, the user-item feedback matrix is completed through emulation. We hypothesise that may explain why the order information is not so informative on the other three datasets.
\vspace{-0.7em}

\subsection{Hyperparameter Sensitivity(RQ5)}
\vspace{-0.2em}
In ROLeR, the major hyperparameter we introduced is the number of nearest neighbors $k$. Thus, we investigate the influence of $k$ on the cumulative rewards. Considering the size and sparsity of the datasets, we test ROLeR in $[5, 10, 15, 20, 30, 40, 50]$ for KuaiRec, $[25, 50, 75, 100, 150, 200]$ for KuaiRand, $[10, 15, 20, 25, 30, 35]$ for Coat, and $[10, 20, 40, 60, 80, 100]$ for Yahoo. 

\begin{figure}[!t]
\centering
\begin{minipage}[b]{0.48\columnwidth}
    \includegraphics[trim=0cm 0cm 0cm 0cm, clip, width=\columnwidth]{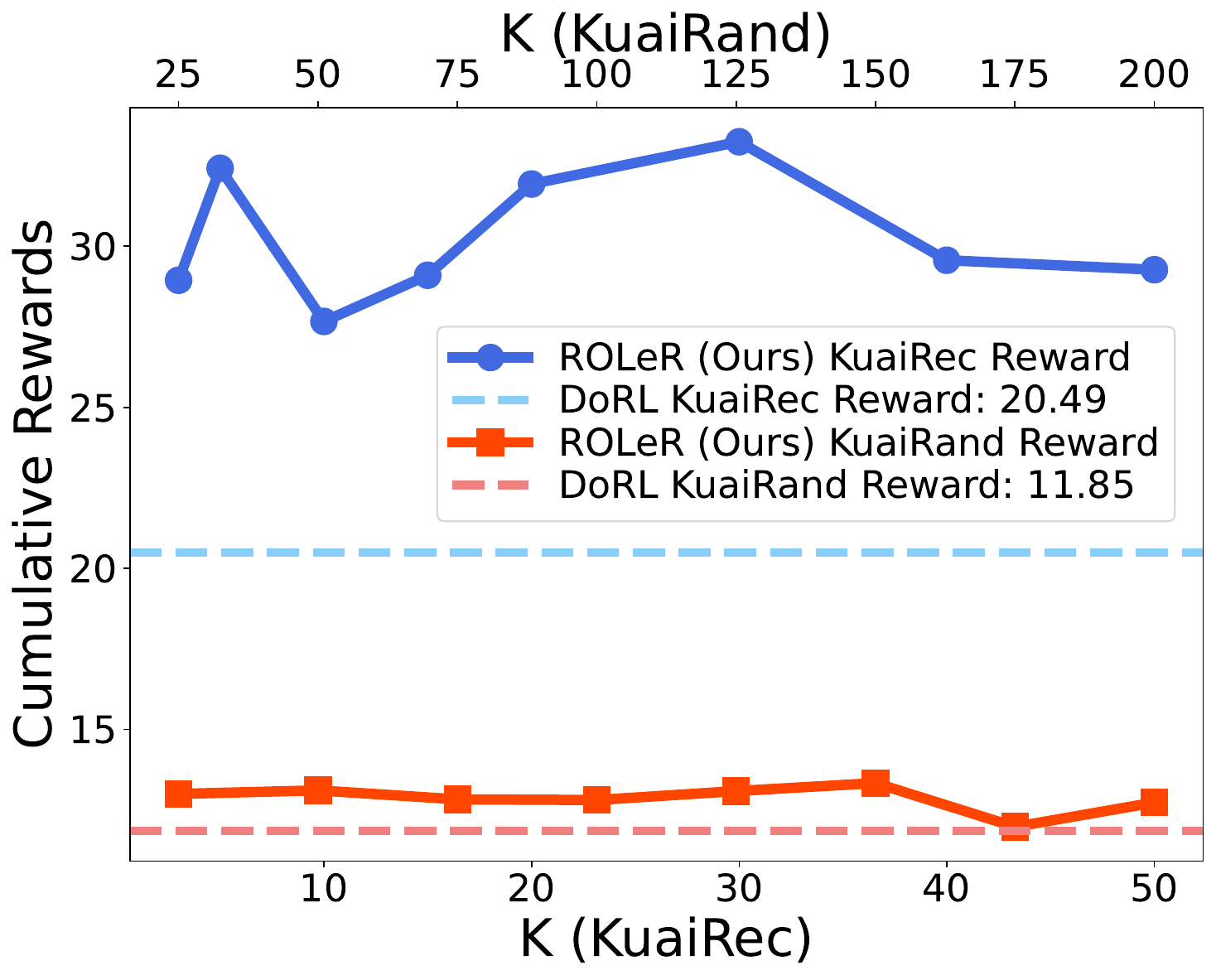}
    \vspace{-2.75em}
\end{minipage}\hfill
\begin{minipage}[b]{0.48\columnwidth}
    \includegraphics[trim=0cm 0cm 0cm 0cm, clip, width=\columnwidth]{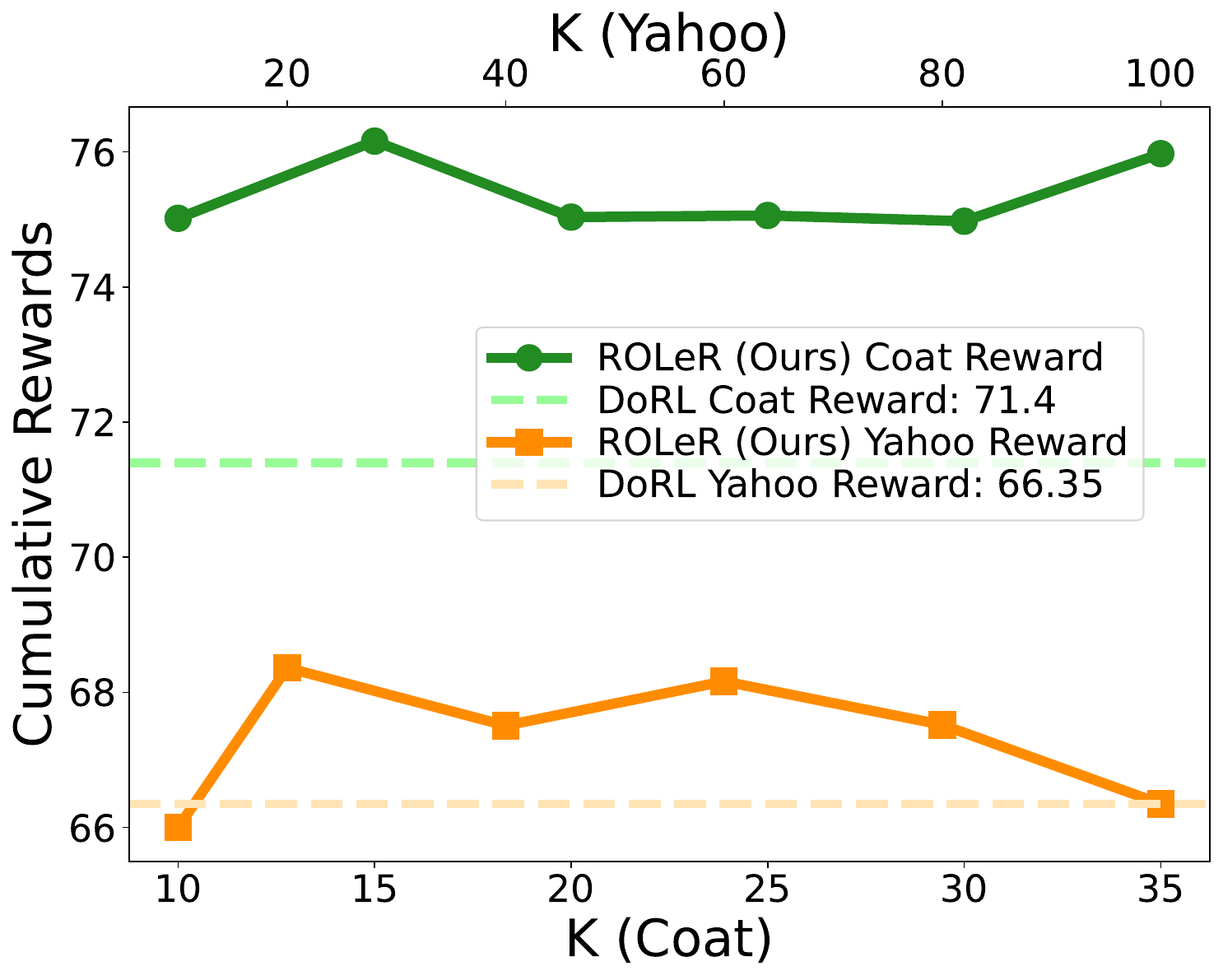}
    \vspace{-2.75em}
    \end{minipage}
\caption{Robustness w.r.t. different $k$s on four datasets.}
\label{fig:robust}
\end{figure}

We draw two line charts to show the relationship between the cumulative rewards and $k$ in Figure~\ref{fig:robust}. The dashed lines in each subplot represent the performance of DORL. We can observe that across a large range of $k$, the cumulative rewards change within a reasonable range, usually less than 2 across all datasets. In addition, ROLeR with diverse $k$ choices constantly surpasses current SOTA on four datasets. It shows the robustness of the proposed ROLeR against its core hyperparameter.
\vspace{-0.7em}

\begin{table}[]
\centering
\small
\caption{The world model impact on KuaiRec}
\vspace{-1em}
\label{tab:world_model}
\scalebox{0.9}{
\begin{tabular}{l|c|l|c}
\toprule
Variants & $R_{tra}$ $\uparrow$ & Variants & $R_{tra}$ $\uparrow$    \\ \midrule\midrule
DORL                             & 20.494 ± 2.670 & ROLeR w. avg \& $a_t$             & 26.006 ± 1.337 \\
DORL w. att                      & 22.499 ± 1.359 & ROLeR w. avg                     & 25.748 ± 1.588  \\
DORL w. $a_G$                    & 22.060 ± 1.098 & ROLeR w. $a_t$                   & 27.646 ± 2.014 \\
DORL w. att \& $a_G$              & 24.431 ± 1.677 & ROLeR                            & 33.246 ± 2.640 \\ \bottomrule
\end{tabular}}
\vspace{-1.8em}
\end{table}

\section{Conclusion}
\vspace{-0.2em}
In this paper, we identify that the reward function prediction is inaccurate in the world model of current model-based offline reinforcement learning for recommender systems. We verify that this inaccuracy can lead to a decrease in the user's long-term satisfaction. Thus, we introduce a non-parametric clustering-based reward shaping method to effectively improve the cumulative reward by enhancing the users' interaction length and single-step reward. In addition, to release the dependence on an ensemble of world models, an in-cluster distance-based uncertainty penalty is proposed to refine the reward shaping by inhibiting risky recommendations. Empirically, we conduct extensive experiments on four challenging datasets. Our method outperforms all baseline methods including the current state-of-the-art.
\vspace{-0.7em}

\section{Acknowledegments}
\vspace{-0.2em}
This work is supported by projects DE200101610, CE200100025 funded by Australian Research Council, and CSIRO’s Science Leader project R-91559.

\bibliographystyle{ACM-Reference-Format}
\bibliography{sample-base}


\begin{thebibliography}{56}


\ifx \showCODEN    \undefined \def \showCODEN     #1{\unskip}     \fi
\ifx \showDOI      \undefined \def \showDOI       #1{#1}\fi
\ifx \showISBNx    \undefined \def \showISBNx     #1{\unskip}     \fi
\ifx \showISBNxiii \undefined \def \showISBNxiii  #1{\unskip}     \fi
\ifx \showISSN     \undefined \def \showISSN      #1{\unskip}     \fi
\ifx \showLCCN     \undefined \def \showLCCN      #1{\unskip}     \fi
\ifx \shownote     \undefined \def \shownote      #1{#1}          \fi
\ifx \showarticletitle \undefined \def \showarticletitle #1{#1}   \fi
\ifx \showURL      \undefined \def \showURL       {\relax}        \fi
\providecommand\bibfield[2]{#2}
\providecommand\bibinfo[2]{#2}
\providecommand\natexlab[1]{#1}
\providecommand\showeprint[2][]{arXiv:#2}

\bibitem[Afsar et~al\mbox{.}(2022)]%
        {afsar2022rsrl}
\bibfield{author}{\bibinfo{person}{M~Mehdi Afsar}, \bibinfo{person}{Trafford Crump}, {and} \bibinfo{person}{Behrouz Far}.} \bibinfo{year}{2022}\natexlab{}.
\newblock \showarticletitle{Reinforcement learning based recommender systems: A survey}.
\newblock \bibinfo{journal}{\emph{Comput. Surveys}} \bibinfo{volume}{55}, \bibinfo{number}{7} (\bibinfo{year}{2022}), \bibinfo{pages}{1--38}.
\newblock


\bibitem[Bobadilla et~al\mbox{.}(2013)]%
        {bobadilla2013rs_survey}
\bibfield{author}{\bibinfo{person}{Jes{\'u}s Bobadilla}, \bibinfo{person}{Fernando Ortega}, \bibinfo{person}{Antonio Hernando}, {and} \bibinfo{person}{Abraham Guti{\'e}rrez}.} \bibinfo{year}{2013}\natexlab{}.
\newblock \showarticletitle{Recommender systems survey}.
\newblock \bibinfo{journal}{\emph{Knowledge-Based Systems}}  \bibinfo{volume}{46} (\bibinfo{year}{2013}), \bibinfo{pages}{109--132}.
\newblock


\bibitem[Burke(2002)]%
        {burke2002hybridRS}
\bibfield{author}{\bibinfo{person}{Robin Burke}.} \bibinfo{year}{2002}\natexlab{}.
\newblock \showarticletitle{Hybrid recommender systems: Survey and experiments}.
\newblock \bibinfo{journal}{\emph{User Modeling and User-Adapted Interaction}}  \bibinfo{volume}{12} (\bibinfo{year}{2002}), \bibinfo{pages}{331--370}.
\newblock


\bibitem[Chen et~al\mbox{.}(2023a)]%
        {chen2023opportunities}
\bibfield{author}{\bibinfo{person}{Xiaocong Chen}, \bibinfo{person}{Siyu Wang}, \bibinfo{person}{Julian McAuley}, \bibinfo{person}{Dietmar Jannach}, {and} \bibinfo{person}{Lina Yao}.} \bibinfo{year}{2023}\natexlab{a}.
\newblock \showarticletitle{On the opportunities and challenges of offline reinforcement learning for recommender systems}.
\newblock \bibinfo{journal}{\emph{ACM Trans. Inf. Syst.}} (\bibinfo{year}{2023}).
\newblock


\bibitem[Chen et~al\mbox{.}(2023b)]%
        {chen2023deep}
\bibfield{author}{\bibinfo{person}{Xiaocong Chen}, \bibinfo{person}{Lina Yao}, \bibinfo{person}{Julian McAuley}, \bibinfo{person}{Guanglin Zhou}, {and} \bibinfo{person}{Xianzhi Wang}.} \bibinfo{year}{2023}\natexlab{b}.
\newblock \showarticletitle{Deep reinforcement learning in recommender systems: A survey and new perspectives}.
\newblock \bibinfo{journal}{\emph{Knowledge-Based Systems}}  \bibinfo{volume}{264} (\bibinfo{year}{2023}), \bibinfo{pages}{110335}.
\newblock


\bibitem[Fujimoto et~al\mbox{.}(2019)]%
        {fujimoto2019bcq}
\bibfield{author}{\bibinfo{person}{Scott Fujimoto}, \bibinfo{person}{David Meger}, {and} \bibinfo{person}{Doina Precup}.} \bibinfo{year}{2019}\natexlab{}.
\newblock \showarticletitle{Off-policy deep reinforcement learning without exploration}. In \bibinfo{booktitle}{\emph{ICML}}. \bibinfo{pages}{2052--2062}.
\newblock


\bibitem[Gao et~al\mbox{.}(2023a)]%
        {gao2023dorl}
\bibfield{author}{\bibinfo{person}{Chongming Gao}, \bibinfo{person}{Kexin Huang}, \bibinfo{person}{Jiawei Chen}, \bibinfo{person}{Yuan Zhang}, \bibinfo{person}{Biao Li}, \bibinfo{person}{Peng Jiang}, \bibinfo{person}{Shiqi Wang}, \bibinfo{person}{Zhong Zhang}, {and} \bibinfo{person}{Xiangnan He}.} \bibinfo{year}{2023}\natexlab{a}.
\newblock \showarticletitle{Alleviating matthew effect of offline reinforcement learning in interactive recommendation}. In \bibinfo{booktitle}{\emph{SIGIR}}.
\newblock


\bibitem[Gao et~al\mbox{.}(2022a)]%
        {gao2022kuairec}
\bibfield{author}{\bibinfo{person}{Chongming Gao}, \bibinfo{person}{Shijun Li}, \bibinfo{person}{Wenqiang Lei}, \bibinfo{person}{Jiawei Chen}, \bibinfo{person}{Biao Li}, \bibinfo{person}{Peng Jiang}, \bibinfo{person}{Xiangnan He}, \bibinfo{person}{Jiaxin Mao}, {and} \bibinfo{person}{Tat-Seng Chua}.} \bibinfo{year}{2022}\natexlab{a}.
\newblock \showarticletitle{KuaiRec: A fully-observed dataset and insights for evaluating recommender systems}. In \bibinfo{booktitle}{\emph{CIKM}}. \bibinfo{pages}{540--550}.
\newblock


\bibitem[Gao et~al\mbox{.}(2022b)]%
        {gao2022kuairand}
\bibfield{author}{\bibinfo{person}{Chongming Gao}, \bibinfo{person}{Shijun Li}, \bibinfo{person}{Yuan Zhang}, \bibinfo{person}{Jiawei Chen}, \bibinfo{person}{Biao Li}, \bibinfo{person}{Wenqiang Lei}, \bibinfo{person}{Peng Jiang}, {and} \bibinfo{person}{Xiangnan He}.} \bibinfo{year}{2022}\natexlab{b}.
\newblock \showarticletitle{KuaiRand: An Unbiased Sequential Recommendation Dataset with Randomly Exposed Videos}. In \bibinfo{booktitle}{\emph{CIKM}}. \bibinfo{pages}{3953--3957}.
\newblock


\bibitem[Gao et~al\mbox{.}(2023b)]%
        {gao2023cirs}
\bibfield{author}{\bibinfo{person}{Chongming Gao}, \bibinfo{person}{Shiqi Wang}, \bibinfo{person}{Shijun Li}, \bibinfo{person}{Jiawei Chen}, \bibinfo{person}{Xiangnan He}, \bibinfo{person}{Wenqiang Lei}, \bibinfo{person}{Biao Li}, \bibinfo{person}{Yuan Zhang}, {and} \bibinfo{person}{Peng Jiang}.} \bibinfo{year}{2023}\natexlab{b}.
\newblock \showarticletitle{CIRS: Bursting filter bubbles by counterfactual interactive recommender system}.
\newblock \bibinfo{journal}{\emph{ACM Trans. Inf. Syst.}} \bibinfo{volume}{42}, \bibinfo{number}{1} (\bibinfo{year}{2023}), \bibinfo{pages}{1--27}.
\newblock


\bibitem[Guo et~al\mbox{.}(2017)]%
        {deepfm}
\bibfield{author}{\bibinfo{person}{Huifeng Guo}, \bibinfo{person}{Ruiming Tang}, \bibinfo{person}{Yunming Ye}, \bibinfo{person}{Zhenguo Li}, {and} \bibinfo{person}{Xiuqiang He}.} \bibinfo{year}{2017}\natexlab{}.
\newblock \showarticletitle{DeepFM: a factorization-machine based neural network for CTR prediction}. In \bibinfo{booktitle}{\emph{IJCAI}}. \bibinfo{pages}{1725–1731}.
\newblock


\bibitem[Haarnoja et~al\mbox{.}(2018)]%
        {haarnoja2018sac}
\bibfield{author}{\bibinfo{person}{Tuomas Haarnoja}, \bibinfo{person}{Aurick Zhou}, \bibinfo{person}{Pieter Abbeel}, {and} \bibinfo{person}{Sergey Levine}.} \bibinfo{year}{2018}\natexlab{}.
\newblock \showarticletitle{Soft actor-critic: Off-policy maximum entropy deep reinforcement learning with a stochastic actor}. In \bibinfo{booktitle}{\emph{ICML}}. \bibinfo{pages}{1861--1870}.
\newblock


\bibitem[Hong et~al\mbox{.}(2020)]%
        {hong2020nonintrusive}
\bibfield{author}{\bibinfo{person}{Daocheng Hong}, \bibinfo{person}{Yang Li}, {and} \bibinfo{person}{Qiwen Dong}.} \bibinfo{year}{2020}\natexlab{}.
\newblock \showarticletitle{Nonintrusive-sensing and reinforcement-learning based adaptive personalized music recommendation}. In \bibinfo{booktitle}{\emph{SIGIR}}. \bibinfo{pages}{1721--1724}.
\newblock


\bibitem[Huang et~al\mbox{.}(2022)]%
        {huang2022state_repr}
\bibfield{author}{\bibinfo{person}{Jin Huang}, \bibinfo{person}{Harrie Oosterhuis}, \bibinfo{person}{Bunyamin Cetinkaya}, \bibinfo{person}{Thijs Rood}, {and} \bibinfo{person}{Maarten de Rijke}.} \bibinfo{year}{2022}\natexlab{}.
\newblock \showarticletitle{State encoders in reinforcement learning for recommendation: A reproducibility study}. In \bibinfo{booktitle}{\emph{SIGIR}}. \bibinfo{pages}{2738--2748}.
\newblock


\bibitem[Janner et~al\mbox{.}(2019)]%
        {janner2019mbpo}
\bibfield{author}{\bibinfo{person}{Michael Janner}, \bibinfo{person}{Justin Fu}, \bibinfo{person}{Marvin Zhang}, {and} \bibinfo{person}{Sergey Levine}.} \bibinfo{year}{2019}\natexlab{}.
\newblock \showarticletitle{When to trust your model: Model-based policy optimization}. In \bibinfo{booktitle}{\emph{NIPS}}. \bibinfo{pages}{12519--12530}.
\newblock


\bibitem[Jeunen and Goethals(2021)]%
        {jeunen2021pessimistic}
\bibfield{author}{\bibinfo{person}{Olivier Jeunen} {and} \bibinfo{person}{Bart Goethals}.} \bibinfo{year}{2021}\natexlab{}.
\newblock \showarticletitle{Pessimistic reward models for off-policy learning in recommendation}. In \bibinfo{booktitle}{\emph{RecSys}}. \bibinfo{pages}{63--74}.
\newblock


\bibitem[Kang and McAuley(2018)]%
        {kang2018sasrec}
\bibfield{author}{\bibinfo{person}{Wang-Cheng Kang} {and} \bibinfo{person}{Julian McAuley}.} \bibinfo{year}{2018}\natexlab{}.
\newblock \showarticletitle{Self-attentive sequential recommendation}. In \bibinfo{booktitle}{\emph{ICDM}}. \bibinfo{pages}{197--206}.
\newblock


\bibitem[Kidambi et~al\mbox{.}(2020)]%
        {kidambi2020morel}
\bibfield{author}{\bibinfo{person}{Rahul Kidambi}, \bibinfo{person}{Aravind Rajeswaran}, \bibinfo{person}{Praneeth Netrapalli}, {and} \bibinfo{person}{Thorsten Joachims}.} \bibinfo{year}{2020}\natexlab{}.
\newblock \showarticletitle{Morel: Model-based offline reinforcement learning}. In \bibinfo{booktitle}{\emph{NIPS}}. \bibinfo{pages}{21810--21823}.
\newblock


\bibitem[Kumar et~al\mbox{.}(2019)]%
        {kumar2019stabilizing}
\bibfield{author}{\bibinfo{person}{Aviral Kumar}, \bibinfo{person}{Justin Fu}, \bibinfo{person}{Matthew Soh}, \bibinfo{person}{George Tucker}, {and} \bibinfo{person}{Sergey Levine}.} \bibinfo{year}{2019}\natexlab{}.
\newblock \showarticletitle{Stabilizing off-policy q-learning via bootstrapping error reduction}. In \bibinfo{booktitle}{\emph{NIPS}}. \bibinfo{pages}{11784--11794}.
\newblock


\bibitem[Kumar et~al\mbox{.}(2020)]%
        {kumar2020cql}
\bibfield{author}{\bibinfo{person}{Aviral Kumar}, \bibinfo{person}{Aurick Zhou}, \bibinfo{person}{George Tucker}, {and} \bibinfo{person}{Sergey Levine}.} \bibinfo{year}{2020}\natexlab{}.
\newblock \showarticletitle{Conservative q-learning for offline reinforcement learning}. In \bibinfo{booktitle}{\emph{NIPS}}. \bibinfo{pages}{1179--1191}.
\newblock


\bibitem[Lai and Robbins(1985)]%
        {lai1985ucb}
\bibfield{author}{\bibinfo{person}{Tze~Leung Lai} {and} \bibinfo{person}{Herbert Robbins}.} \bibinfo{year}{1985}\natexlab{}.
\newblock \showarticletitle{Asymptotically efficient adaptive allocation rules}.
\newblock \bibinfo{journal}{\emph{Advances in Applied Mathematics}} \bibinfo{volume}{6}, \bibinfo{number}{1} (\bibinfo{year}{1985}), \bibinfo{pages}{4--22}.
\newblock


\bibitem[Lei and Li(2019)]%
        {user_profile12}
\bibfield{author}{\bibinfo{person}{Yu Lei} {and} \bibinfo{person}{Wenjie Li}.} \bibinfo{year}{2019}\natexlab{}.
\newblock \showarticletitle{Interactive recommendation with user-specific deep reinforcement learning}.
\newblock \bibinfo{journal}{\emph{ACM Trans. Knowl. Discov. Data}} \bibinfo{volume}{13}, \bibinfo{number}{6} (\bibinfo{year}{2019}), \bibinfo{pages}{1--15}.
\newblock


\bibitem[Lei et~al\mbox{.}(2020)]%
        {GCQN_RS}
\bibfield{author}{\bibinfo{person}{Yu Lei}, \bibinfo{person}{Hongbin Pei}, \bibinfo{person}{Hanqi Yan}, {and} \bibinfo{person}{Wenjie Li}.} \bibinfo{year}{2020}\natexlab{}.
\newblock \showarticletitle{Reinforcement learning based recommendation with graph convolutional q-network}. In \bibinfo{booktitle}{\emph{SIGIR}}. \bibinfo{pages}{1757--1760}.
\newblock


\bibitem[Lei et~al\mbox{.}(2019)]%
        {lei2019social}
\bibfield{author}{\bibinfo{person}{Yu Lei}, \bibinfo{person}{Zhitao Wang}, \bibinfo{person}{Wenjie Li}, {and} \bibinfo{person}{Hongbin Pei}.} \bibinfo{year}{2019}\natexlab{}.
\newblock \showarticletitle{Social attentive deep q-network for recommendation}. In \bibinfo{booktitle}{\emph{SIGIR}}. \bibinfo{pages}{1189--1192}.
\newblock


\bibitem[Levine et~al\mbox{.}(2020)]%
        {levine2020offline}
\bibfield{author}{\bibinfo{person}{Sergey Levine}, \bibinfo{person}{Aviral Kumar}, \bibinfo{person}{George Tucker}, {and} \bibinfo{person}{Justin Fu}.} \bibinfo{year}{2020}\natexlab{}.
\newblock \showarticletitle{Offline Reinforcement Learning: Tutorial, Review, and Perspectives on Open Problems}.
\newblock \bibinfo{journal}{\emph{CoRR}}  \bibinfo{volume}{abs/2005.01643} (\bibinfo{year}{2020}).
\newblock


\bibitem[Liu et~al\mbox{.}(2021)]%
        {liu2021interaction}
\bibfield{author}{\bibinfo{person}{Ping Liu}, \bibinfo{person}{Karthik Shivaram}, \bibinfo{person}{Aron Culotta}, \bibinfo{person}{Matthew~A Shapiro}, {and} \bibinfo{person}{Mustafa Bilgic}.} \bibinfo{year}{2021}\natexlab{}.
\newblock \showarticletitle{The interaction between political typology and filter bubbles in news recommendation algorithms}. In \bibinfo{booktitle}{\emph{Web Conf}}. \bibinfo{pages}{3791--3801}.
\newblock


\bibitem[Lowe et~al\mbox{.}(2017)]%
        {lowe2017maddpg}
\bibfield{author}{\bibinfo{person}{Ryan Lowe}, \bibinfo{person}{Yi Wu}, \bibinfo{person}{Aviv Tamar}, \bibinfo{person}{Jean Harb}, \bibinfo{person}{OpenAI Pieter~Abbeel}, {and} \bibinfo{person}{Igor Mordatch}.} \bibinfo{year}{2017}\natexlab{}.
\newblock \showarticletitle{Multi-agent actor-critic for mixed cooperative-competitive environments}. In \bibinfo{booktitle}{\emph{NIPS}}. \bibinfo{pages}{6382--6393}.
\newblock


\bibitem[Marlin and Zemel(2009)]%
        {marlin2009yahoo}
\bibfield{author}{\bibinfo{person}{Benjamin~M Marlin} {and} \bibinfo{person}{Richard~S Zemel}.} \bibinfo{year}{2009}\natexlab{}.
\newblock \showarticletitle{Collaborative prediction and ranking with non-random missing data}. In \bibinfo{booktitle}{\emph{RecSys}}. \bibinfo{pages}{5--12}.
\newblock


\bibitem[Mnih et~al\mbox{.}(2016)]%
        {mnih2016a2c}
\bibfield{author}{\bibinfo{person}{Volodymyr Mnih}, \bibinfo{person}{Adria~Puigdomenech Badia}, \bibinfo{person}{Mehdi Mirza}, \bibinfo{person}{Alex Graves}, \bibinfo{person}{Timothy Lillicrap}, \bibinfo{person}{Tim Harley}, \bibinfo{person}{David Silver}, {and} \bibinfo{person}{Koray Kavukcuoglu}.} \bibinfo{year}{2016}\natexlab{}.
\newblock \showarticletitle{Asynchronous methods for deep reinforcement learning}. In \bibinfo{booktitle}{\emph{ICML}}. \bibinfo{pages}{1928--1937}.
\newblock


\bibitem[Pazis and Parr(2013)]%
        {pazis2013pac}
\bibfield{author}{\bibinfo{person}{Jason Pazis} {and} \bibinfo{person}{Ronald Parr}.} \bibinfo{year}{2013}\natexlab{}.
\newblock \showarticletitle{PAC optimal exploration in continuous space Markov decision processes}. In \bibinfo{booktitle}{\emph{AAAI}}. \bibinfo{pages}{774--781}.
\newblock


\bibitem[Prudencio et~al\mbox{.}(2023)]%
        {prudencio2023survey}
\bibfield{author}{\bibinfo{person}{Rafael~Figueiredo Prudencio}, \bibinfo{person}{Marcos~ROA Maximo}, {and} \bibinfo{person}{Esther~Luna Colombini}.} \bibinfo{year}{2023}\natexlab{}.
\newblock \showarticletitle{A survey on offline reinforcement learning: Taxonomy, review, and open problems}.
\newblock \bibinfo{journal}{\emph{IEEE Trans. Neural Netw. Learn. Syst}} (\bibinfo{year}{2023}).
\newblock


\bibitem[Qiu et~al\mbox{.}(2021)]%
        {qiu2021memory}
\bibfield{author}{\bibinfo{person}{Ruihong Qiu}, \bibinfo{person}{Zi Huang}, {and} \bibinfo{person}{Hongzhi Yin}.} \bibinfo{year}{2021}\natexlab{}.
\newblock \showarticletitle{Memory augmented multi-instance contrastive predictive coding for sequential recommendation}. In \bibinfo{booktitle}{\emph{ICDM}}. \bibinfo{pages}{519--528}.
\newblock


\bibitem[Qiu et~al\mbox{.}(2022)]%
        {qiu2022contrastive}
\bibfield{author}{\bibinfo{person}{Ruihong Qiu}, \bibinfo{person}{Zi Huang}, \bibinfo{person}{Hongzhi Yin}, {and} \bibinfo{person}{Zijian Wang}.} \bibinfo{year}{2022}\natexlab{}.
\newblock \showarticletitle{Contrastive learning for representation degeneration problem in sequential recommendation}. In \bibinfo{booktitle}{\emph{WSDM}}. \bibinfo{pages}{813--823}.
\newblock


\bibitem[Qiu et~al\mbox{.}(2019)]%
        {qiu2019rethinking}
\bibfield{author}{\bibinfo{person}{Ruihong Qiu}, \bibinfo{person}{Jingjing Li}, \bibinfo{person}{Zi Huang}, {and} \bibinfo{person}{Hongzhi Yin}.} \bibinfo{year}{2019}\natexlab{}.
\newblock \showarticletitle{Rethinking the item order in session-based recommendation with graph neural networks}. In \bibinfo{booktitle}{\emph{CIKM}}. \bibinfo{pages}{579--588}.
\newblock


\bibitem[Qiu et~al\mbox{.}(2020)]%
        {qiu2020gag}
\bibfield{author}{\bibinfo{person}{Ruihong Qiu}, \bibinfo{person}{Hongzhi Yin}, \bibinfo{person}{Zi Huang}, {and} \bibinfo{person}{Tong Chen}.} \bibinfo{year}{2020}\natexlab{}.
\newblock \showarticletitle{Gag: Global attributed graph neural network for streaming session-based recommendation}. In \bibinfo{booktitle}{\emph{SIGIR}}. \bibinfo{pages}{669--678}.
\newblock


\bibitem[Schnabel et~al\mbox{.}(2016)]%
        {schnabel2016coat}
\bibfield{author}{\bibinfo{person}{Tobias Schnabel}, \bibinfo{person}{Adith Swaminathan}, \bibinfo{person}{Ashudeep Singh}, \bibinfo{person}{Navin Chandak}, {and} \bibinfo{person}{Thorsten Joachims}.} \bibinfo{year}{2016}\natexlab{}.
\newblock \showarticletitle{Recommendations as treatments: Debiasing learning and evaluation}. In \bibinfo{booktitle}{\emph{ICML}}. \bibinfo{pages}{1670--1679}.
\newblock


\bibitem[Schulman et~al\mbox{.}(2017)]%
        {schulman2017ppo}
\bibfield{author}{\bibinfo{person}{John Schulman}, \bibinfo{person}{Filip Wolski}, \bibinfo{person}{Prafulla Dhariwal}, \bibinfo{person}{Alec Radford}, {and} \bibinfo{person}{Oleg Klimov}.} \bibinfo{year}{2017}\natexlab{}.
\newblock \showarticletitle{Proximal Policy Optimization Algorithms}.
\newblock \bibinfo{journal}{\emph{CoRR}}  \bibinfo{volume}{abs/1707.06347} (\bibinfo{year}{2017}).
\newblock


\bibitem[Sutton and Barto(2018)]%
        {sutton2018reinforcement}
\bibfield{author}{\bibinfo{person}{Richard~S Sutton} {and} \bibinfo{person}{Andrew~G Barto}.} \bibinfo{year}{2018}\natexlab{}.
\newblock \bibinfo{booktitle}{\emph{Reinforcement learning: An introduction}}.
\newblock \bibinfo{publisher}{MIT press}.
\newblock


\bibitem[Swaminathan and Joachims(2015)]%
        {swaminathan2015ips}
\bibfield{author}{\bibinfo{person}{Adith Swaminathan} {and} \bibinfo{person}{Thorsten Joachims}.} \bibinfo{year}{2015}\natexlab{}.
\newblock \showarticletitle{Counterfactual risk minimization: Learning from logged bandit feedback}. In \bibinfo{booktitle}{\emph{ICML}}. \bibinfo{pages}{814--823}.
\newblock


\bibitem[Van~Hasselt et~al\mbox{.}(2016)]%
        {van2016ddqn}
\bibfield{author}{\bibinfo{person}{Hado Van~Hasselt}, \bibinfo{person}{Arthur Guez}, {and} \bibinfo{person}{David Silver}.} \bibinfo{year}{2016}\natexlab{}.
\newblock \showarticletitle{Deep reinforcement learning with double q-learning}. In \bibinfo{booktitle}{\emph{AAAI}}. \bibinfo{pages}{2094--2100}.
\newblock


\bibitem[Vaswani et~al\mbox{.}(2017)]%
        {vaswani2017attention}
\bibfield{author}{\bibinfo{person}{Ashish Vaswani}, \bibinfo{person}{Noam Shazeer}, \bibinfo{person}{Niki Parmar}, \bibinfo{person}{Jakob Uszkoreit}, \bibinfo{person}{Llion Jones}, \bibinfo{person}{Aidan~N Gomez}, \bibinfo{person}{{\L}ukasz Kaiser}, {and} \bibinfo{person}{Illia Polosukhin}.} \bibinfo{year}{2017}\natexlab{}.
\newblock \showarticletitle{Attention is all you need}. In \bibinfo{booktitle}{\emph{NIPS}}. \bibinfo{pages}{6000--6010}.
\newblock


\bibitem[Wang and Mu(2020)]%
        {wang2020aspect}
\bibfield{author}{\bibinfo{person}{Hengliang Wang} {and} \bibinfo{person}{Kedian Mu}.} \bibinfo{year}{2020}\natexlab{}.
\newblock \showarticletitle{Aspect-Level Attributed Network Embedding via Variational Graph Neural Networks}. In \bibinfo{booktitle}{\emph{DASFAA}}. \bibinfo{pages}{398--414}.
\newblock


\bibitem[Wang et~al\mbox{.}(2020a)]%
        {wang2020statistical}
\bibfield{author}{\bibinfo{person}{Ruosong Wang}, \bibinfo{person}{Dean~P. Foster}, {and} \bibinfo{person}{Sham~M. Kakade}.} \bibinfo{year}{2020}\natexlab{a}.
\newblock \showarticletitle{What are the Statistical Limits of Offline {RL} with Linear Function Approximation?}
\newblock \bibinfo{journal}{\emph{CoRR}}  \bibinfo{volume}{abs/2010.11895} (\bibinfo{year}{2020}).
\newblock


\bibitem[Wang et~al\mbox{.}(2022)]%
        {wang2022surrogate}
\bibfield{author}{\bibinfo{person}{Yuyan Wang}, \bibinfo{person}{Mohit Sharma}, \bibinfo{person}{Can Xu}, \bibinfo{person}{Sriraj Badam}, \bibinfo{person}{Qian Sun}, \bibinfo{person}{Lee Richardson}, \bibinfo{person}{Lisa Chung}, \bibinfo{person}{Ed~H Chi}, {and} \bibinfo{person}{Minmin Chen}.} \bibinfo{year}{2022}\natexlab{}.
\newblock \showarticletitle{Surrogate for long-term user experience in recommender systems}. In \bibinfo{booktitle}{\emph{SIGKDD}}. \bibinfo{pages}{4100--4109}.
\newblock


\bibitem[Wang et~al\mbox{.}(2020b)]%
        {wang2020crr}
\bibfield{author}{\bibinfo{person}{Ziyu Wang}, \bibinfo{person}{Alexander Novikov}, \bibinfo{person}{Konrad Zolna}, \bibinfo{person}{Josh~S Merel}, \bibinfo{person}{Jost~Tobias Springenberg}, \bibinfo{person}{Scott~E Reed}, \bibinfo{person}{Bobak Shahriari}, \bibinfo{person}{Noah Siegel}, \bibinfo{person}{Caglar Gulcehre}, \bibinfo{person}{Nicolas Heess}, {et~al\mbox{.}}} \bibinfo{year}{2020}\natexlab{b}.
\newblock \showarticletitle{Critic regularized regression}. In \bibinfo{booktitle}{\emph{NIPS}}. \bibinfo{pages}{7768--7778}.
\newblock


\bibitem[Xiao et~al\mbox{.}(2020)]%
        {xiao2020deep}
\bibfield{author}{\bibinfo{person}{Yilin Xiao}, \bibinfo{person}{Liang Xiao}, \bibinfo{person}{Xiaozhen Lu}, \bibinfo{person}{Hailu Zhang}, \bibinfo{person}{Shui Yu}, {and} \bibinfo{person}{H~Vincent Poor}.} \bibinfo{year}{2020}\natexlab{}.
\newblock \showarticletitle{Deep-reinforcement-learning-based user profile perturbation for privacy-aware recommendation}.
\newblock \bibinfo{journal}{\emph{IEEE IoT-J}} (\bibinfo{year}{2020}).
\newblock


\bibitem[Xin et~al\mbox{.}(2020)]%
        {xin2020sqn}
\bibfield{author}{\bibinfo{person}{Xin Xin}, \bibinfo{person}{Alexandros Karatzoglou}, \bibinfo{person}{Ioannis Arapakis}, {and} \bibinfo{person}{Joemon~M Jose}.} \bibinfo{year}{2020}\natexlab{}.
\newblock \showarticletitle{Self-supervised reinforcement learning for recommender systems}. In \bibinfo{booktitle}{\emph{SIGIR}}. \bibinfo{pages}{931--940}.
\newblock


\bibitem[Xu et~al\mbox{.}(2022)]%
        {xu2022policy}
\bibfield{author}{\bibinfo{person}{Haoran Xu}, \bibinfo{person}{Li Jiang}, \bibinfo{person}{Li Jianxiong}, {and} \bibinfo{person}{Xianyuan Zhan}.} \bibinfo{year}{2022}\natexlab{}.
\newblock \showarticletitle{A policy-guided imitation approach for offline reinforcement learning}. In \bibinfo{booktitle}{\emph{NIPS}}. \bibinfo{pages}{4085--4098}.
\newblock


\bibitem[Yu et~al\mbox{.}(2021)]%
        {yu2021combo}
\bibfield{author}{\bibinfo{person}{Tianhe Yu}, \bibinfo{person}{Aviral Kumar}, \bibinfo{person}{Rafael Rafailov}, \bibinfo{person}{Aravind Rajeswaran}, \bibinfo{person}{Sergey Levine}, {and} \bibinfo{person}{Chelsea Finn}.} \bibinfo{year}{2021}\natexlab{}.
\newblock \showarticletitle{Combo: Conservative offline model-based policy optimization}. In \bibinfo{booktitle}{\emph{NIPS}}. \bibinfo{pages}{28954--28967}.
\newblock


\bibitem[Yu et~al\mbox{.}(2020)]%
        {yu2020mopo}
\bibfield{author}{\bibinfo{person}{Tianhe Yu}, \bibinfo{person}{Garrett Thomas}, \bibinfo{person}{Lantao Yu}, \bibinfo{person}{Stefano Ermon}, \bibinfo{person}{James~Y Zou}, \bibinfo{person}{Sergey Levine}, \bibinfo{person}{Chelsea Finn}, {and} \bibinfo{person}{Tengyu Ma}.} \bibinfo{year}{2020}\natexlab{}.
\newblock \showarticletitle{Mopo: Model-based offline policy optimization}. In \bibinfo{booktitle}{\emph{NIPS}}. \bibinfo{pages}{14129--14142}.
\newblock


\bibitem[Zhang et~al\mbox{.}(2019)]%
        {zhang2019deep}
\bibfield{author}{\bibinfo{person}{Shuai Zhang}, \bibinfo{person}{Lina Yao}, \bibinfo{person}{Aixin Sun}, {and} \bibinfo{person}{Yi Tay}.} \bibinfo{year}{2019}\natexlab{}.
\newblock \showarticletitle{Deep learning based recommender system: A survey and new perspectives}.
\newblock \bibinfo{journal}{\emph{ACM computing surveys}} \bibinfo{volume}{52}, \bibinfo{number}{1} (\bibinfo{year}{2019}), \bibinfo{pages}{1--38}.
\newblock


\bibitem[Zhao et~al\mbox{.}(2018)]%
        {zhao2018deep}
\bibfield{author}{\bibinfo{person}{Xiangyu Zhao}, \bibinfo{person}{Long Xia}, \bibinfo{person}{Liang Zhang}, \bibinfo{person}{Zhuoye Ding}, \bibinfo{person}{Dawei Yin}, {and} \bibinfo{person}{Jiliang Tang}.} \bibinfo{year}{2018}\natexlab{}.
\newblock \showarticletitle{Deep reinforcement learning for page-wise recommendations}. In \bibinfo{booktitle}{\emph{RecSys}}. \bibinfo{pages}{95--103}.
\newblock


\bibitem[Zhao et~al\mbox{.}(2020)]%
        {zhao2020deepchain}
\bibfield{author}{\bibinfo{person}{Xiangyu Zhao}, \bibinfo{person}{Long Xia}, \bibinfo{person}{Lixin Zou}, \bibinfo{person}{Hui Liu}, \bibinfo{person}{Dawei Yin}, {and} \bibinfo{person}{Jiliang Tang}.} \bibinfo{year}{2020}\natexlab{}.
\newblock \showarticletitle{Whole-chain recommendations}. In \bibinfo{booktitle}{\emph{CIKM}}. \bibinfo{pages}{1883--1891}.
\newblock


\bibitem[Zheng et~al\mbox{.}(2018)]%
        {zheng2018drn}
\bibfield{author}{\bibinfo{person}{Guanjie Zheng}, \bibinfo{person}{Fuzheng Zhang}, \bibinfo{person}{Zihan Zheng}, \bibinfo{person}{Yang Xiang}, \bibinfo{person}{Nicholas~Jing Yuan}, \bibinfo{person}{Xing Xie}, {and} \bibinfo{person}{Zhenhui Li}.} \bibinfo{year}{2018}\natexlab{}.
\newblock \showarticletitle{DRN: A deep reinforcement learning framework for news recommendation}. In \bibinfo{booktitle}{\emph{Web Conf}}. \bibinfo{pages}{167--176}.
\newblock


\bibitem[Zheng et~al\mbox{.}(2021)]%
        {zheng2021dgcn}
\bibfield{author}{\bibinfo{person}{Yu Zheng}, \bibinfo{person}{Chen Gao}, \bibinfo{person}{Liang Chen}, \bibinfo{person}{Depeng Jin}, {and} \bibinfo{person}{Yong Li}.} \bibinfo{year}{2021}\natexlab{}.
\newblock \showarticletitle{DGCN: Diversified recommendation with graph convolutional networks}. In \bibinfo{booktitle}{\emph{Web Conf}}. \bibinfo{pages}{401--412}.
\newblock


\bibitem[Zou et~al\mbox{.}(2019)]%
        {user_profile1}
\bibfield{author}{\bibinfo{person}{Lixin Zou}, \bibinfo{person}{Long Xia}, \bibinfo{person}{Zhuoye Ding}, \bibinfo{person}{Jiaxing Song}, \bibinfo{person}{Weidong Liu}, {and} \bibinfo{person}{Dawei Yin}.} \bibinfo{year}{2019}\natexlab{}.
\newblock \showarticletitle{Reinforcement learning to optimize long-term user engagement in recommender systems}. In \bibinfo{booktitle}{\emph{SIGKDD}}. \bibinfo{pages}{2810--2818}.
\newblock


\end{thebibliography}

\end{document}